\documentclass{emulateapj}
\usepackage{apjfonts}

\newcommand{\kpc}{\>{\rm kpc}} \newcommand{\Mpc}{\>{\rm Mpc}}
\newcommand{\kms}{\>{\rm km}\,{\rm s}^{-1}} \newcommand{\Msun}{\>{\rm
    M_{\odot}}}
\def\be{\begin{equation}} \def\ee{\end{equation}}
 \def\fig#1{Fig.~\ref{fig:#1}}
\def\equ#1{Eq.~(\ref{eq:#1})} \def\tab#1{Table ~\ref{tb:#1}}

\begin{document}

\title{THE ANGULAR MOMENTUM DISTRIBUTION OF GAS AND DARK MATTER IN
  GALACTIC HALOS}

\author{Sanjib Sharma} \affil{Department of Physics University of
  Arizona, Tucson, AZ 85721, USA \\ Astrophysikalisches Institut
  Potsdam, 14482 Potsdam, Germany} \email{ssharma@aip.de} \and
\author{Matthias Steinmetz\altaffilmark{1}} \affil{Astrophysikalisches
  Institut
  Potsdam, 14482 Potsdam, Germany \\
  Steward Observatory, University of Arizona, 933 N Cherry Ave,
  Tucson, AZ 85721, USA} \email{msteinmetz@aip.de}
\altaffiltext{1}{David and Lucile Packard Fellow}

\begin{abstract}
  
  We report results of a series of non radiative N-body/SPH
  simulations in a $\Lambda CDM$ cosmology, designed to study the
  growth of angular momentum in galaxy systems. A sample of 41 halos
  of differing mass and environment were selected from a cosmological
  N-body simulation of size $32.5h^{-1}$Mpc, and re-simulated at
  higher resolution with the tree-SPH code GADGET.
  
  We find that the spin of the baryonic component correlates well with
  the spin of the dark matter, but there is a misalignment of
  typically $20^{\circ}$ between these two components.  The spin of
  the baryonic component is also on average larger than that of the
  dark matter component and we find this effect to be more pronounced
  at lower redshifts. A significant fraction $f$ of gas has negative
  angular momentum and this fraction is found to increase with
  redshift. This trend can be explained as a result of increasing
  thermalization of the virializing gas with decreasing redshift.  We
  describe a toy model in which the tangential velocities of particles
  are smeared by Gaussian random motions.  This model is successful in
  explaining some of the global angular momentum properties, in
  particular the anti-correlation of $f$ with the spin parameter
  $\lambda$ , and the shape of the angular momentum distributions.
  
  We investigate in detail the angular momentum distributions (AMDs)
  of the gas and the dark matter components of the halo. We compare
  and contrast various techniques to determine the AMDs. We show that
  broadening of velocity dispersions is unsuitable for making
  comparisons between gas and dark matter AMDs because the shape of
  the broadened AMDs is predominantly determined by the dispersion and
  is insensitive to the underlying non broadened AMD. In order to
  bring both gas and dark matter to the same footing, we smooth the
  angular momentum of the particles over a fixed number of neighbors.
  The AMDs obtained by this method have a smooth and extended
  truncation as compared to earlier methods. We find that an
  analytical function in which the differential distribution of
  specific angular momentum $j$ is given by $P(j) =
  \frac{1}{j_{d}^{\alpha}\Gamma(\alpha)}(j)^{\alpha-1}e^{-j/j_{d}} $,
  where $j_d=j_{tot}/\alpha$, can be used to describe a wide variety
  of profiles, with just one parameter $\alpha$. The distribution of
  the shape parameter $\alpha$ for both gas and dark matter follows
  roughly a log-normal distribution. The mean and standard deviation
  of $log(\alpha)$ for gas is $-0.04$ and $0.11$ respectively. About
  $90-95 \%$ of halos have $\alpha<1.3$, while exponential disks in
  NFW halos would require $1.3<\alpha<1.6$.  This implies that a
  typical halo in simulations has an excess of low angular momentum
  material as compared to that of observed exponential disks, a result
  which is consistent with the findings of earlier works.  $\alpha$
  for gas is correlated with that for dark matter (DM) but they have a
  significant scatter $<\alpha_{Gas}/\alpha_{DM}>=1.09 \pm 0.2$.
  $\alpha_{Gas}$ is also biased towards slightly higher values
  compared to $\alpha_{DM}$. The angular momentum in halos is also
  found to have a significant spatial asymmetry with the asymmetry
  being more pronounced for dark matter.
\end{abstract}


\section{Introduction} \label{sec:intro}

Disk galaxies are rotationally supported systems and their structural
properties are intimately linked to their angular momentum
distribution.  The standard picture of formation of disk galaxies is
that the density perturbations grow due to gravitational instability
and end up forming virialized systems of dark matter and gas. The gas
cools and collapses towards the center \citep[]{1978MNRAS.183..341W}.
The gas has angular momentum which it acquires due to tidal
interactions.  This can be quantified in terms of a dimensionless spin
parameter $\lambda=J|E|^{1/2}/(GM^{5/2})$
\citep[]{1969ApJ...155..393P} which has a value of about $0.05$
\citep[]{1979MNRAS.186..133E,1987ApJ...319..575B,1995MNRAS...272..570B}.
The angular momentum of the gas is conserved during the collapse
resulting in the formation of a centrifugally supported disk, whose
size is consistent with that of observed disk galaxies
\citep[]{1980MNRAS.193..189F}.

Based on the initial density profiles ($\rho=\rho(r)$) and angular
momentum distributions ($m=m(j)$) or ($M_{<j}=M_{<j}(j)$) the final
surface density of disks can be determined.
\citet{1997ApJ...482..659D} derived the surface density of the disks
assuming the halos to be uniform spheres in solid body rotation.  From
numerical simulations it is now known that density profiles of dark
matter in halos follow a universal profile as described by
\citet[NFW]{1996ApJ...462..563N,1997ApJ...490..493N}. Using these
realistic profiles and assuming the final surface density of disks to
be exponential \citet{1998MNRAS.295..319M} investigated various
properties of disks like rotation curves, disk scale lengths and so
on.  They also addressed the issue of stability of disks. However,
angular momentum distributions are required to more realistically
model the surface density of disk galaxies.  \citet[henceforth
B2001]{2001ApJ...555..240B} have reported that in CDM simulations, the
DM halos obey a universal angular momentum distribution of the form

\begin{equation}
  \label{Mj-j}
  \frac{M(<j)}{M_v} = \frac{\mu j}{j_0 + j}
\end{equation}
where $j$ is the specific angular momentum and $M(<j)$ is mass with
specific angular momentum less than $j$. The shape parameter $\mu$ has
a log-normal distribution and the $90\%$ range is given by
$1.05<\mu<2.0$.  Assuming that the angular momentum profile of gas is
identical to that of dark matter, B2001 calculated the surface density
profiles of the resulting disks, and found that (for the range of
$\mu$ given above) the resulting disks are too centrally concentrated
compared to exponential profiles. In addition, detailed hydrodynamical
simulations of gas collapse in hierarchical structure formation
scenario exhibit the so-called ``angular momentum catastrophe'': the
gas component looses its angular momentum due to dynamical friction
and ends up forming disks that are far too concentrated
\citep[]{1991ApJ...380..320N,
  1994MNRAS.267..401N,1997ApJ...478...13N,1999ApJ...513..555S}.

Recently \citet{2002ApJ...576...21V} (henceforth vB2002) has tested
the assumption, that the AMDs of gas and DM are similar, by directly
measuring the AMDs of gas in proto galaxies from hydrodynamical
simulations at $z=3$. They find the spin parameter distribution of gas
and dark matter to be identical in spite of the angular momentum
vectors of gas and dark matter being misaligned by $\sim 35^\circ$. In
order to compare the AMDs of gas and DM within halos, they broaden the
velocities of gas to account for the microscopic random motions of gas
atoms.  The broadened profiles are found to be very similar to that of
DM, though, as we are arguing later in this paper, this result is
dominated by the dispersion hiding away the effect of the actual
profiles.
 
vB2002 also demonstrated that a considerable amount of gas (between 5
and 50 percent) have negative angular momentum.  Assuming that the gas
with negative angular momentum combines with that of positive angular
momentum to form a non rotating bulge, and the remaining positive
angular momentum material ends up forming a disk, they found the
surface density profiles of the resulting disks to closely follow an
exponential profile. However the galaxies end up with a large $B/D$
ratio, and with a minimum $B/D$ of 0.1 the question of how to form
bulge-less dwarf and LSB galaxies is still unanswered.  The problem of
an excess of low AM material gets transferred into a problem of
excessive bulge formation.

To investigate these issues in more detail and to see if these
properties have any evolution with redshift we perform a series of
N-Body/SPH simulations of selected halos till z=0.  Special care is
taken to have high number of particles in the final virialized halos
so as to measure the angular momentum accurately. The details of the
simulation and methods of analysis are described in Sec-2 . Results
related to global angular momentum parameters and their evolution with
redshift are presented in Sec-3.  In Sec-4 we present a toy model to
explain some of these findings.  In Sec-5 and 6 we analyze the angular
momentum distributions.

\section{Method}

\subsection{Simulation}
The cosmology adopted in the simulation is the so called concordance
model of $\Lambda CDM$ cosmology, in agreement with recent WMAP and
SDSS results \citep[]{2003ApJS..148..175S,2003ApJ...586L...1M}.  The
parameters adopted are $\Omega_{\lambda}=0.7$ ,$\Omega_{m}=0.3$ and
$\Omega_{b}=0.02235 h^{-2}$ and a Hubble constant of $65 \kms
\Mpc^{-1}$ . The power spectrum parameter determining the amplitude of
mass fluctuations in a sphere of radius $8 h^{-1} \Mpc$ was set to
$0.9$ and shape parameter $\Gamma$ was set to $0.2$. An $AP^3M$ code
was used to evolve $128^3$ dark matter particles in a $32.5 h^{-1}
\Mpc$ cube from $z=24$ to $z=0$ using 2000 equal steps in expansion
factor.  At $z=0$, 41 halos were selected with circular velocities
ranging from $64 \kms$ to $310 \kms$ and masses ranging from $1.3
\times 10^{11} \Msun $ to $ 1.5 \times 10^{13} \Msun$ .  These halos
were then re-simulated at higher resolution with more dark matter
particles and also with an equal number of gas particles.  The
simulations were performed from $z=50$ to $z=0$. These re-simulations
were done using the code GADGET \citep[]{2001NewA....6...79S}.  The
gas particles were given an initial temperature of $100 K$ at $z=50$
and an artificial temperature floor of 100K was kept during the
simulation. The number of particles (of each kind ) within the virial
radius ranges from $8000-80000$. A gravitational softening of $2 \kpc
\ h^{-1}$ (physical) was used. The integration was performed in
comoving co-ordinates.

\subsection{Halo Identification}
We adopt the method of vB2002 to identify the center of mass of a
virialized region.  We start with the densest gas particle as a guess
for the center of mass and iteratively increase the radius till the
average mass density enclosed by a spherical region is $\Delta_v$
times the mean matter density at that redshift. $\Delta_v$ is
approximated by \citep[]{1998ApJ...495...80B} $ \Delta_v \simeq
(18\pi^2 +82x+-39x^2)/(1+x)$,where $x=\Omega_{m}(z)-1$.  We re-center
the particles within the virial radius in velocities and position and
then recalculate the virial radius based on this new center of mass.
We repeat this process until the distance between the center of mass
before and after the calculation of virial radius is less than 0.1
percent of the virial radius.

\subsection{Angular momentum distributions}
As discussed by vB2002, there are two kinds of velocities for
particles in the simulations, the actual microscopic velocity $\bf
{v}$ of individual particles and the mean streaming velocity ${\bf u}$
at any location ${\bf x}$ . The actual microscopic velocity $\bf {v}$
is given by equation ${\bf v = u+ w}$, where ${\bf w}$ is the
particles random motion. For collisionless dark matter particles,
which interact only through gravity, the velocity given by simulations
is ${\bf v}$, whereas collisional gas (SPH) particles, the velocity
given by simulations is ${\bf u}$, the information about the random
motion is incorporated into the internal energy per unit mass $U$. If
$\sigma$ is one dimensional velocity dispersion of the particle,its
temperature $T$ is given by
\begin{equation}
  \label{eq:u-kT}
  U=\frac{3}{2}\sigma^2=\frac{3}{2}\frac{kT}{\mu}
\end{equation}
where $\mu$ is the mean molecular weight of gas. In order to compare
the kinematical properties of the gaseous or dark matter component we
either need to broaden the velocities of gas particles by using
\equ{u-kT} (we label these by superscript ${ \bf t}$; denoting the
actual motion ) or smoothen the velocities of dark matter particles by
an appropriate smoothening length (we label these by superscript ${\bf
  s}$; denoting the streaming motion of the fluid).

The total angular momentum of gas or dark matter is given by
\begin{equation}
  \label{eq:J-def}
  {\bf J}_{gas,DM} =  \sum_{i=1}^{N_{gas},N_{DM}}  m_i {\bf r_i}  
  \times {\bf v_i}
\end{equation}
where ${\bf r_i}$ and ${\bf v_i}$ are the radius and velocity vectors
respectively of a particle $i$, in a co-ordinate system in which both
the position and the velocity of the center of mass of the entire halo
(DM + gas) is zero.  For the spin parameter, we use the modified
definition of B2001.
\begin{eqnarray}
  \label{eq:lambda-jrv}
  \lambda_{gas,DM}= \frac{ |{\bf j}_{gas,DM}| }{\sqrt{2} R_{vir} V_{vir}}
\end{eqnarray}
where $ j_{gas}$ and $ j_{DM}$ are the mean specific angular momentum
of gas and dark matter respectively, and $V_{vir} =
\sqrt{G(M_{gas}+M_{DM})/{R_{vir}}}$ is the circular velocity at virial
radius $R_{vir}$.  The misalignment $\theta$ between the angular
momentum vectors of gas and dark matter ${\bf J_{gas}}$ and ${ \bf
  J_{DM}}$ is given by
\begin{equation}
  \label{eq:theta-def}
  cos \theta=( \frac{{\bf J_{gas} \cdot J_{DM}}}{\bf{ |J_{gas}| |J_{DM}|}} )
\end{equation}

We make a co-ordinate transformation such that the z-axis is aligned
with the total angular momentum vector. For gas particles the z-axis
is aligned along ${\bf J_{gas}}$ while for dark matter z-axis is
aligned along ${\bf J_{DM}}$. The $z$ component of specific angular
momentum $j_z$ is then measured (henceforth we drop the subscript and
denote it by $j$).  The fraction of mass with $j<0$ is labeled as
$f_{gas}$ and $f_{DM}$ for gas and dark matter respectively. The
differential angular momentum distribution (AMD) is the fraction of
mass $P(j)$ with specific angular momentum between $j$ to $j+dj$ i.e.
$\int_{-\infty}^{\infty}P(j)dj=1$.  We define a parameter $l$ which is
related to $j$ by
\begin{equation}
  l=\frac{j}{\sqrt{2}V_{vir}R_{vir}}
\end{equation}
(similar to definition in vB2002 except for a factor of $\sqrt{2}$).
The above definition for $l$ implies that
\begin{equation}
  \int_{-1}^{1}lP(l)dl=\lambda
\end{equation}

The total specific angular momentum of a halo is denoted by $j_{tot}$
and this is related to $\lambda$ by
$j_{tot}=\lambda\sqrt{2}R_{vir}V_{vir}$, see \equ{lambda-jrv}. The
shape and extent of profiles depend on $j_{tot}$.  With $s=j/j_{tot}$
,as proposed in \citet[ henceforth BBS01]{2001MNRAS.326.1205V}, we
obtain
\begin{equation}
  \int_{-\infty}^{\infty}sP(s)ds=1
\end{equation}

In most of our analysis we neglect the negative tail and define the
distributions for the positive tail only and normalize with respect to
it. The range of $s$ or $j$ in that case is from $0$ to $\infty$ and
for halos with negative tails the quantities like $j_{tot}$ also need
to be recalculated for the positive tail only.  The cumulative angular
momentum distribution $P(<j)$ is the fraction of mass with AM less
than equal to $j$. It is also defined only for the positive tail, and
is normalized with respect to it.

\section{Results}

\subsection{Global angular momentum parameters at z=3}
\begin{figure}
  \epsscale{1.1} \plotone{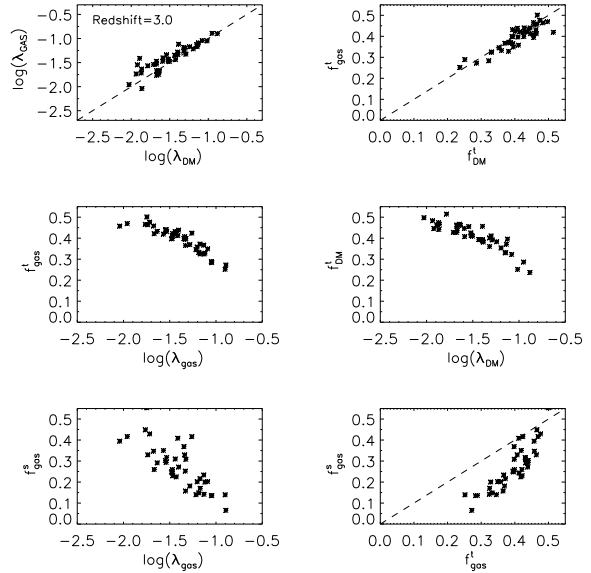}
  \caption{A comparison of various global angular momentum
    parameters at $z=3$ for both gas and dark matter. The parameters
    compared are the spin parameter $\lambda$ and the fraction with
    negative angular momentum $f$. A comparison with Fig-2 of vB2002
    shows that at $z=3$ the results of our simulations are in good
    agreement with the analysis of vB2002.
    \label{fig:properties2z3_crop}}
\end{figure}
As a first test we compare the results of our simulations against the
analysis of vB2002, who compared various global angular momentum
parameters using a sample of 378 halos at a redshift of z=3 ( see
Fig-2 in vB2002). Our results for a sample of 41 halos at the same
redshift, are shown in \fig{properties2z3_crop}. We find them to be in
good agreement.

The first plot on upper left compares $\lambda_{gas}$ against
$\lambda_{DM}$. $\lambda$ of gas and DM are well correlated with a
Spearman rank coefficient of $r_s=0.88$, but show significant scatter.
The mean of $\lambda_{gas}/\lambda_{DM}$ is $1.27$ with a standard
deviation of $0.40$.  This suggests that statistically the angular
momentum acquired by gas and dark matter is similar but on a
one-by-one comparison the spin parameters can be quite different.  The
fraction of matter with negative angular momentum for gas and dark
matter denoted by $f_{gas}^{t}$ and $f_{DM}^{t}$, are also well
correlated with Spearman rank coefficient of 0.81 and
$<f_{DM}^{t}/f_{gas}^{t}> = 0.96 \pm 0.07$ (As described in Sec-2.3
the superscript $t$ denotes inclusion of microscopic random motions
while superscript $s$ stands for streaming motions only).  In the
middle panels the strong anti-correlation of $f_{gas}^{t}$ and
$f_{DM}^{t}$ with their respective spin parameters can be seen. In the
lower left panel $f_{gas}^{s}$ is also found to be anti-correlated
with $\lambda_{gas}$.  If we assume that the random motions are
responsible for the negative angular momentum and that the ordered
motion contributes to the $\lambda$, then this result is easy to
understand.  The more the ordered motion the less will be the effect
of random motions.  The plot of $f_{gas}^{s}$ vs $\lambda_{gas}$ is
found to have more scatter than that of $f_{gas}^{t}$ vs
$\lambda_{gas}$ which has broadened velocities.  We will follow up on
this though in more detail in section 4.

In the lower left plot $f_{gas}^{s}$ is shown against $f_{gas}^{t}$.
$f_{gas}^{s}$ is less than $f_{gas}^{t}$ as expected ( because
$f_{gas}^{t}$ has more random motion) and the difference $f_{gas}^{s}-
f_{gas}^{t}$ is found to increase at lower $f$ values or at lower
$\lambda$, respectively.  A comparison of some of our results with
vB2002 is given in \tab{global}.

\begin{deluxetable}{lll}
  \tablecaption{ Global angular momentum parameters } \tablehead{
    \colhead{ } & \colhead{ vB2002 } & \colhead{This paper}}
  \startdata
  $\bar{\lambda}_{gas}$                          & 0.039                      & 0.040            \\
  $\bar{\lambda}_{DM}$                          & 0.040                      & 0.030            \\
  $\sigma_{gas}$                                    & 0.57                        & 0.62                       \\
  $\sigma_{DM}$                                    & 0.56                        & 0.81                       \\
  $<\lambda_{gas} - \lambda_{DM}>$   & $-0.001\pm 0.020$ & $0.006\pm 0.009$ \\
  $<f_{gas}^{t}/f_{DM}^{t}>$                   & $0.97\pm 0.10$      & $0.96\pm 0.07$   \\
  $<\theta>$                                           & 36.2                        & 18.9             \\
  \enddata \tablecomments{ The properties shown above are for halos at
    $z=3$, alongside are results from vB2002 also at same redshift.
    The results reported in this paper are in good agreement with
    vB2002}
  \label{tb:global}
\end{deluxetable}

\subsection{Redshift dependence of angular momentum parameters}
\subsubsection{Distribution of $\lambda_{DM}$ and $\lambda_{gas}$}
It has been suggested by numerous studies that the distribution of
$\lambda$ can be described by a log-normal distribution
\begin{equation}
  p(\lambda){\rm d} \lambda = {1 \over \sqrt{2 \pi} \sigma_{\lambda}}
  \exp\biggl(- {{\rm ln}^2(\lambda/\bar{\lambda}) \over 2
    \sigma^2_{\lambda}}\biggr) {{\rm d} \lambda \over \lambda}.
  \label{eq:spindistr}
\end{equation}
\citet{2002ApJ...576...21V} have found that the gas and dark matter
have very similar distribution of spin parameters. They found
$\bar{\lambda}_{gas}=0.039$ and $\bar{\lambda}_{DM}=0.040$.

\begin{figure}
  \epsscale{1.1} \plotone{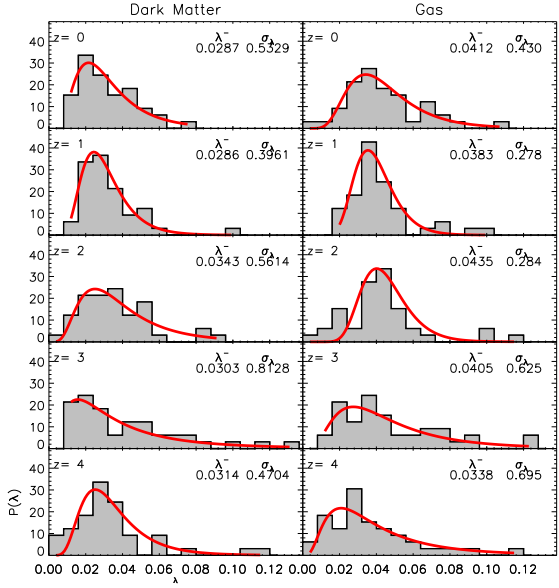}
  \caption{Distribution of spin parameter $\lambda$ for dark matter
    (left) and gas (right) for a sample of 41 halos for redshifts
    ranging from 4 to 0.  The solid line is the best fit log-normal
    distribution as given by \equ{spindistr}. The fit parameters
    $\bar{\lambda}$ and $\sigma_{\lambda}$ are also labeled on each of
    the plots.
\label{fig:lambda-dist}}
\end{figure}

\begin{figure}
  \epsscale{1.1} \plotone{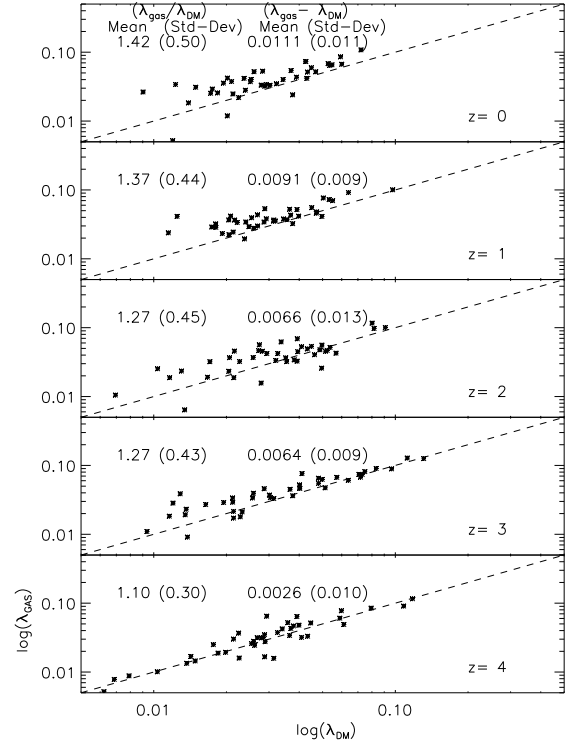}
  \caption{$\lambda_{Gas}$ vs $\lambda_{DM}$ for various redshifts.
    $<\lambda_{gas}/\lambda_{DM}>$ and $<\lambda_{gas}-\lambda_{DM}>$
    along with their standard deviations are shown on each plot.
    \label{fig:lambdag-lambdad}}
\end{figure}

In \fig{lambda-dist} we have plotted the distribution of
$\lambda_{gas}$ and $\lambda_{DM}$ for redshifts from $0$ to $4$ . It
is observed that $\bar{\lambda}_{DM}$ fluctuates between $0.029$ to
$0.034$ while $\bar{\lambda}_{gas}$ seems to increase from 0.034 to
0.041 with decrease in redshift.  The fact that $\lambda_{gas}$ is
higher than $\lambda_{DM}$ can be seen more clearly in
\fig{lambdag-lambdad} where $\lambda_{gas}$ is plotted against
$\lambda_{DM}$.  $<\lambda_{gas}/\lambda_{DM}>$ is greater than 1 and
increases as z decreases.

\subsubsection{Distribution of Misalignment Angle $\theta$}
\begin{figure}
  \epsscale{1.1} \plotone{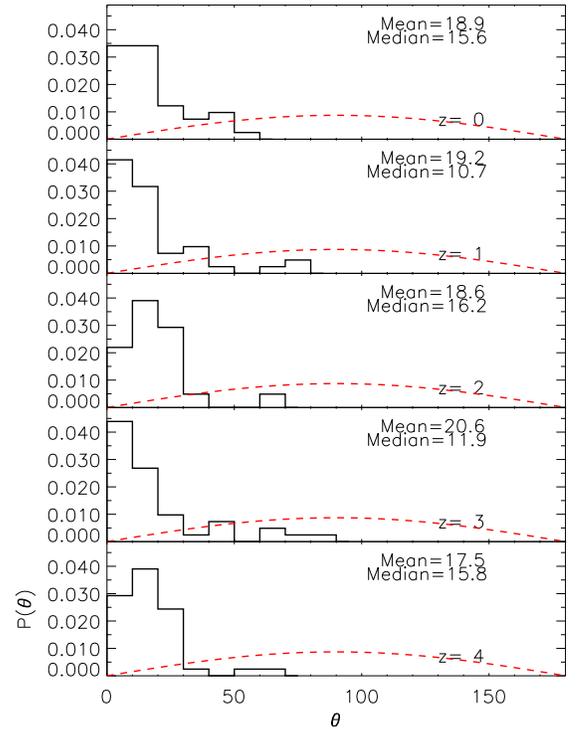}
\caption{Distribution of misalignment angle $\theta$ for various
  redshifts.  The dashed line is the expected distribution if the
  angle is completely random or uncorrelated.
\label{fig:misaln-dist}}
\end{figure}

\begin{figure*}
  \epsscale{1.15} \plotone{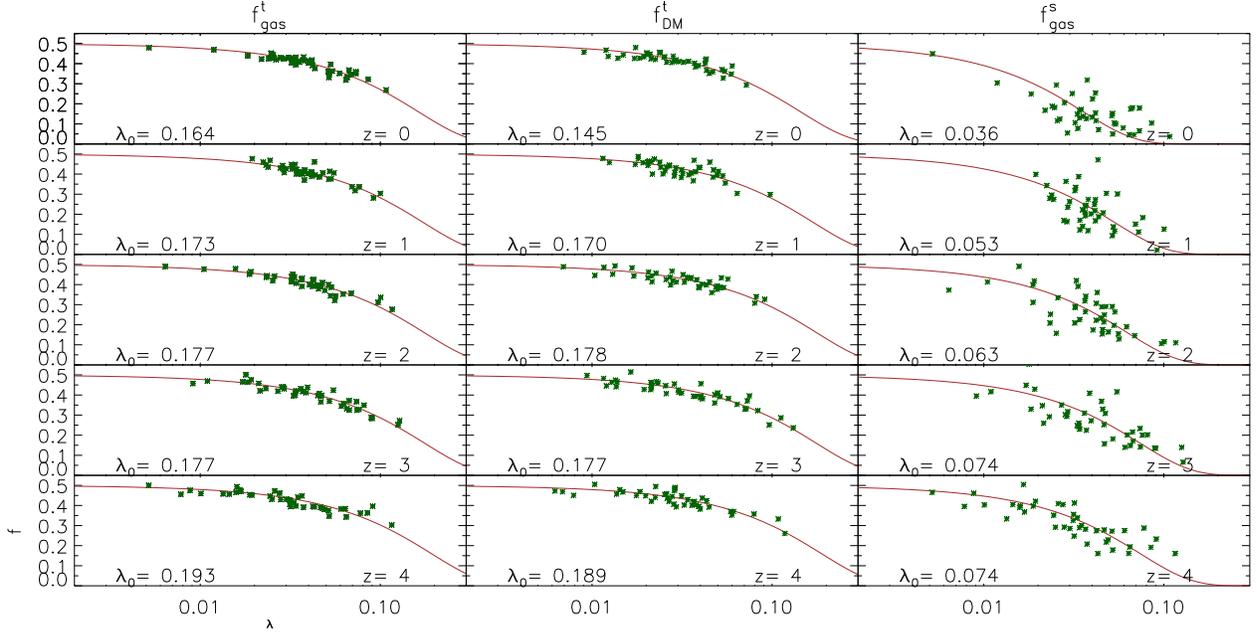}
  \caption{Counter-rotating fraction $f$ as a function of spin
    parameter $\lambda$ for various redshifts.  Superscript $t$ stands
    for analysis done with velocities which incorporate random motions
    and $s$ stands for analysis done with streaming velocities. The
    solid lines are the fits given by the functional form in
    \equ{f-lambda}. The best value of parameter $\lambda_0$ is
    indicated on each plot.
    \label{fig:gl-prop5}}
\end{figure*}
In \fig{misaln-dist} we show the misalignment angle between the total
angular momentum vectors of dark matter and gas for various redshifts.
The mean and median values are also shown in each panel.  The
distribution does not show any significant change with redshift.  The
mean value $<\theta>$ is around $20$ degrees.

\subsubsection{Correlation of $f$ with $\lambda$}
As mentioned in \fig{properties2z3_crop}, the fraction of matter with
negative angular momentum $f$ is anti-correlated with $\lambda$.  We
found that the $f$ vs $\lambda$ distribution can be well fit by a
function of the form
\begin{equation}
  \label{eq:f-lambda}
  f  =  1-I_{g}( \frac{ \lambda}{ \lambda_{0}}) \sim 0.5 e^{-\lambda/\lambda_0}
\end{equation}
where $I_{g}(x)= \int_{-\infty}^{x} \frac{e^{-t^2/(2\sigma^2)}}
{\sqrt{2\pi}\sigma} dt$ is a Gaussian integral.  The results of the
applied fit are shown on each plot in \fig{gl-prop5}.  The
anti-correlation can be described by a single parameter $\lambda_0$.
For the actual motion of both gas and DM, it is found that $\lambda_0$
is around $0.17$ and does not seem to change much except for a
decrease in its value at z=0. For streaming motion of gas the
parameter $\lambda_0$ decreases consistently with decrease in
redshift, thus hinting to a systematical change in the distribution of
$f_{gas}$ with redshift.  Furthermore, the scatter in $f$ vs $\lambda$
plots corresponding to streaming motion of gas ( column 3
\fig{gl-prop5}) is quite large specially at lower redshifts as
compared to the plots for broadened motions (column 1 and 2
\fig{gl-prop5}).

\subsection{Variation of $<f_{gas}>$ with thermalization of gas}
\begin{figure}
  \epsscale{1.1} \plotone{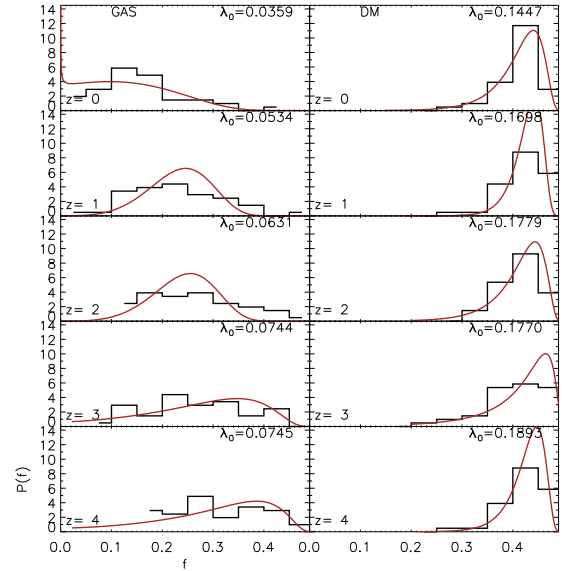}
  \caption{Distribution of counter-rotating fraction $f$ for gas and
    dark matter for a sample of 41 halos.  The distribution is almost
    independent of redshift for dark matter while for gas $f$ is lower
    at lower redshifts.  The solid line is the distribution as
    predicted by \equ{Pf}.
\label{fig:predict_fdist}}
\end{figure}
\begin{figure}
  \plotone{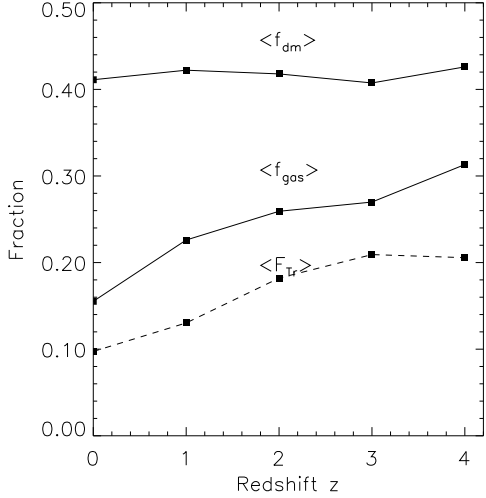} \epsscale{1.1}
  \caption{Counter-rotating fractions $<f_{gas}>$ , $<f_{dm}>$ and the
    fraction of energy for gas in the form of translational motion
    $<F_{Tr}>$, as a function of redshift. The average is taken over a
    sample of 41 halos. At lower redshifts the gas gets more and more
    thermalized, its kinetic energy getting converted into thermal,
    and this decreases $<f>$.
    \label{fig:thermal1}}
\end{figure}

In \fig{predict_fdist} the distribution of $f$ is plotted for various
redshifts.  It can be seen that the distribution for dark matter does
not seem to change with redshift, while for gas it shifts towards
smaller fractions of negative angular momentum material at lower
redshifts.  For DM $<f> \sim 0.40$ while for gas $<f>$ decreases from
0.33 to 0.15 monotonically with decrease in redshift (\fig{thermal1}).
The kinetic energy of gas particles in SPH simulation, is a
combination of translational energy and internal (thermal) energy.
The ratio $F_{Tr}=KE_{Tr}/(KE_{Tr} + KE_{Th} )$ gives an estimate of
thermalization. The lower the $F_{Tr}$ the greater the thermalization.
\fig{thermal1} shows that $<F_{Tr}>$ just like $<f_{gas}>$, also
decreases monotonically with decrease of redshift.  The correlation of
$<F_{Tr}>$ with $<f_{gas}>$ is easy to understand.

The dispersion in velocity of particles gives rise to particles with,
negative angular momentum. For very small velocity dispersion of
particles the fraction with negative angular momentum $f \sim 0 $. If
velocity dispersion is very large or AM is very small, then $f \sim
0.5 $.  The SPH particles only have macroscopic flow velocities, the
thermal energy is incorporated into U. For DM particles all the energy
is in the form of velocities of particles. So $f_{gas}$ is always less
than $f_{dm}$.  At higher redshift there are more mergers and the gas
is more turbulent.  As the redshift decreases, the gas undergoes
relaxation and kinetic energy of gas gets converted into thermal
energy. The velocity dispersion of gas decreases resulting in a
decrement of $f_{gas}$.  For dark matter $F_{Tr}=1$ and $<f>$ does not
show any considerable evolution.

\subsection{Effect of numerical resolution}
\begin{figure}
  \epsscale{1.1} \plotone{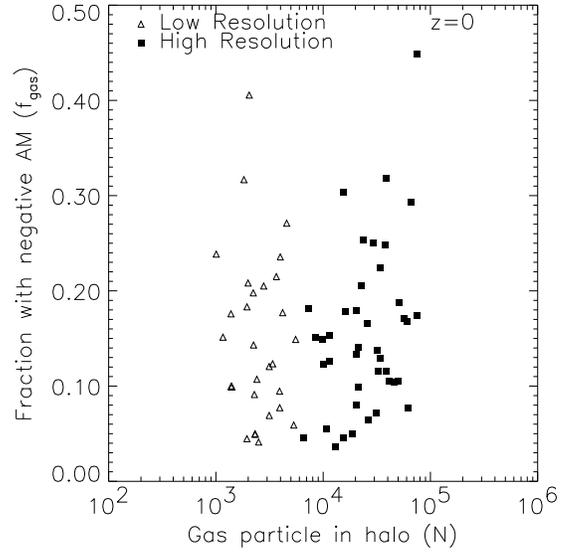}
  \caption{  $f_{gas}$ vs $N_{gas}$ for halos at
    redshift of $z=0$ simulated at two different resolutions.  The
    fraction $f_{gas}$ does not show any apparent correlation with
    number of gas particles in the virialized halo indicating that
    this result is not affected by numerical resolution. The same set
    of halos re-simulated at $1/8$ times the original resolution also
    do not show any systematic shift.
    \label{fig:f_Vs_N}}
\end{figure}
\begin{figure}
  \epsscale{1.1} \plotone{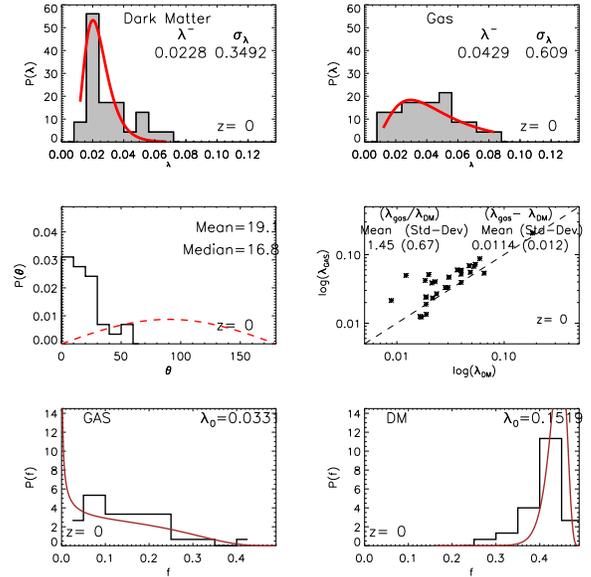}
  \caption{ Various properties of halos and their distributions at
    $z=0$ from low resolution simulations (1/8 times the normal
    resolution).  A comparison with \fig{lambda-dist},
    \fig{lambdag-lambdad}, \fig{misaln-dist} , \fig{predict_fdist}
    which are the corresponding plots with high resolution simulations
    shows no difference.
    \label{fig:lowres_prop1}}
\end{figure}

To check whether the numerical resolution has an effect on $f_{gas}$
we plot $f_{gas}$ versus $N$, the number of gas particles in the halo
(filled squares in \fig{f_Vs_N}) and find no correlation. To check the
effect of numerical resolution in more detail $30$ halos were
re-simulated at a lower resolution with $1/8$ times the original
number of particles (open triangles in \fig{f_Vs_N}).  The values of
$f$ do not show any systematic trend with change of resolution.  Halos
simulated in lower resolution have distributions of $\lambda$ ,$f$ and
misalignment angle $\theta$ (\fig{lowres_prop1}) identical to the
distributions found in high resolution simulations.  So the results
shown above are robust to the effect of numerical resolution.

\subsection{Evolution of gas particles having negative angular momentum} 
\begin{figure}
  \epsscale{1.1} \plotone{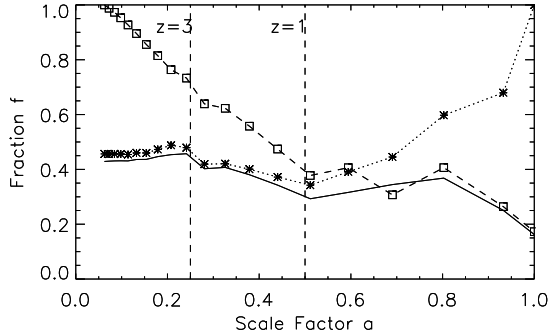} \figcaption{ The evolution of
    fraction $f$ of particles with negative angular momentum with
    scale factor $a$ for various sub-components of gas.  Solid line -
    $f_{global}$ vs $a$, Dotted line(with stars) - $f_{z=0}$ vs $a$
    and Dashed line(with squares) - $f_{z=15}$ vs $a$.  $f_{global}$
    is the fraction of gas with negative angular momentum for the
    whole halo. $f_{z=0}$ is the corresponding fraction out of a
    subset of particles that had negative angular momentum at $z=0$.
    $f_{z=15}$ is the fraction out of a subset of particles that had
    negative angular momentum at $z=15$.
\label{fig:fvsa_a}}
\end{figure}

We investigate whether the gas that ends up with negative angular
momentum also had negative angular momentum in the past.  We identify
a region in past that ends up in the virialized halo at z=0 by simply
tracking halo particles at z=0 back in time. We re-center them, and
then calculate their angular momentum. The fraction of particles with
negative AM at any given stage during the evolution of this Lagrangian
volume is a function of redshift so we denote it by $f_{global}(a)$.
We identify a subset of particles that were counter rotating at $z=0$,
and denote them by subscript $z=0$. The fraction of particles out of
this subset that are counter rotating at any given instant is denoted
by $f_{z=0}(a)$.  Similarly $f_{z=15}(a)$ is the fraction of counter
rotating particles out of a subset of particles that were counter
rotating at z=15.  If the particles with negative angular momentum at
$z=0$ were also counter rotating at an earlier epoch, then they would
have $f_{z=0}(a) \sim 1 $ independent of redshift.  Similarly if
counter rotating particles at $z=15$ also had negative angular
momentum later during their evolution then $f_{z=15}(a) \sim 1$. On
the other hand if $f_{z=15}$ or $f_{z=0} \sim f_{global}$ then the
particles are just a random subset drawn from the original halo and
have no relation with their past or future, respectively.

In \fig{fvsa_a} we have plotted $f_{global}$, $f_{z=0}$ and$f_{z=15}$
as a function of scale factor $a$ of the universe for a randomly
selected halo.  $f_{z=0}$ and $f_{z=15}$ are both close to 1 at their
respective ends but by $z=1$ both fractions have dropped to
$f_{global}$ and continue to remain so in respective directions. This
suggests that particles that are counter rotating at a given instant
will not necessarily counter rotate at a later epoch in future but
will instead get mixed up randomly with the remaining portion of the
halo.

\section{A toy model: Gaussian smearing of ordered velocities (GSOV)}
To get a better intuitive understanding of some of the results
presented so far, in particular the anti-correlation of $f$ with
$\lambda$, we present here a toy model that describes the motion of
the particles in the halo.  We take a co-ordinate system with the $z$
axis pointing along the direction of the total angular momentum
vector. The actual velocity ${\bf u}$ of a particle consists of the
ordered motion $v_o \hat{\phi} $ (which is motion in circular orbits
at a constant speed $v_o$), superimposed by random motions ${\bf
  v_{\sigma}}$.  Each component of random motion ${\bf v_{\sigma}}$ is
drawn out from a Gaussian distribution with a dispersion $\sigma$.
The random motions are assumed to be isotropic. ${\bf u}$ can be
written as
\begin{eqnarray}
{\bf u} & = & v_{o} \hat{\phi} + v_{\sigma_x } {\bf i} +v_{\sigma_y }  
{\bf j} +v_{\sigma_z } {\bf k}  \nonumber \\
   & = & ( v_{o}+ v_{\sigma_{\phi}}) \hat{\phi} + v_{\sigma_{\rho}} 
\hat{\rho} + v_{\sigma_{z}} {\bf k} 
\label{eq:gauss_vel}
\end{eqnarray}
According to this model the histogram of $u_{\phi}$ in a halo is given
by
\begin{eqnarray}
P(u_{\phi})du_{\phi} & = & \frac{1}{\sqrt{2\pi} \sigma_{\phi}}e^{-(u_{\phi}-v_{o})^2/(2\sigma_{\phi}^2)} du_{\phi} 
\label{eq:gauss_func1}
\end{eqnarray}

\begin{figure}
  \epsscale{1.1} \plotone{{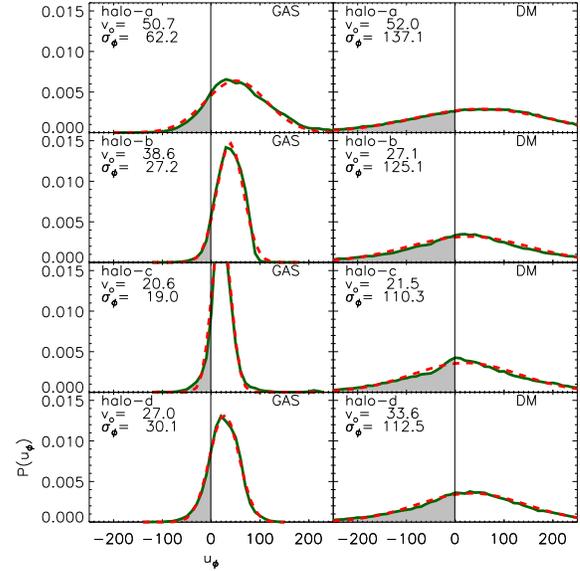}} \figcaption{ A plot showing
    distribution of $u_{\phi}$ for gas and DM for four different
    halos. Solid line: the histogram for particles in simulation,
    Dashed line: A Gaussian fit to the above histogram . The shaded
    region corresponds to particles with negative angular momentum.
\label{fig:gauss_vel1}}
\end{figure}

In \fig{gauss_vel1} we have plotted the histogram of $u_{\phi}$ for
various halos, for both dark matter and gas. For most halos the
$u_{\phi}$ distribution of gas particles can be well described by a
Gaussian. Some halos are biased either to right or left, and some show
two peaks which can described by two Gaussian functions. A closer
examination reveals that these peculiarities are associated with
substructure of halos. For dark matter $\sigma$ is large and this
washes out any peculiarities that may have been present, consequently
no significant deviation from the single Gaussian structure can be
observed.  Sometimes a slightly sharper peak compared to a Gaussian
distribution can be seen indicative of a velocity dispersion that is
not strictly isothermal.

\begin{figure}
  \epsscale{1.1} \plotone{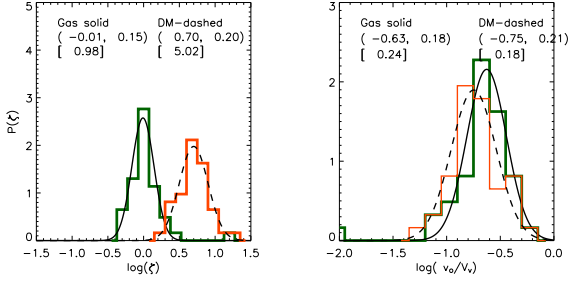} \figcaption{ Distribution of
    parameter $\zeta=\frac{\sigma}{v_{o}}$ and $\frac{v_{o}}{V_{vir}}$
    for a sample of 41 halos at z=0.  Distributions are roughly
    log-normal. Gaussian fits are applied and the fit parameters are
    indicated in upper left and right corners.  The position of the
    peak in normal units is shown below the fit parameters.
\label{fig:zeta_dist}}
\end{figure}

In terms of this model the motion within a halo can be described by
two parameters $v_o$ and $\zeta$ where
$\zeta=\frac{\sigma_{\phi}}{v_o}$.  $\zeta$ is a measure of random
motion relative to ordered motion.  We calculate $v_{o}$ and
$\sigma_{\phi}$ for simulated halos by fitting a Gaussian profile as
given by \equ{gauss_func1} to the histogram of $u_{\phi}$.
\fig{zeta_dist} shows the distribution of $\zeta$ and $v_o/V_v$ for
both gas and DM at $z=0$.  $log(\zeta)$ can be fit by a Gaussian both
for DM and gas.  For gas $\bar{\zeta} \sim 1$ while for DM
$\bar{\zeta} \sim 5$.

\begin{figure}
  \epsscale{1.1} \plotone{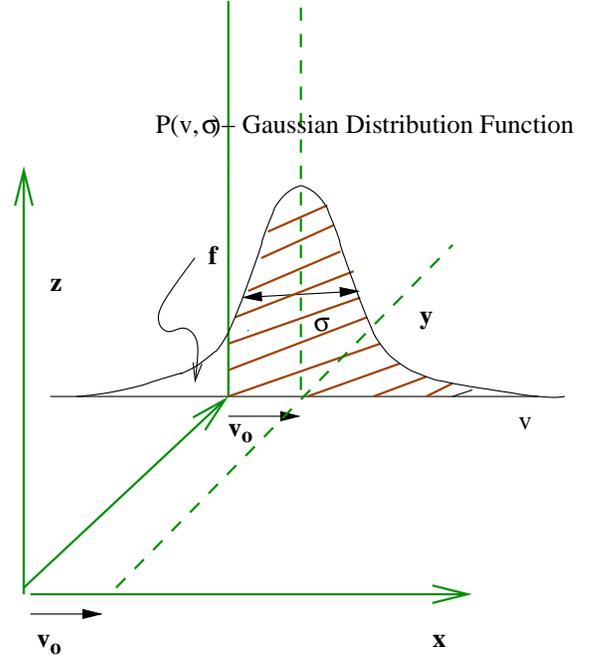} \figcaption{Dispersion of ordered
    velocity $v_{o}$ by a Gaussian with width $\sigma$.
    $P(v,\sigma)=\frac{1}{\sqrt{2\pi}
      \sigma_{\phi}}e^{-v^2/(2\sigma_{\phi}^2)}$ is the Gaussian
    distribution function. A particle has negative angular momentum if
    $v_{\sigma_{\phi}} < -v_{o}$, this is represented by the region
    under the Gaussian curve which is not shaded.
\label{fig:gauss_vel}}
\end{figure}

If the above model of Gaussian smearing is a realistic representation
then the fraction of matter with negative angular momentum is simply
the probability that $u_{\phi}$ is less than zero, i.e.\ 
$v_{\sigma_{\phi}}$ is less than $-v_{o}$ (\fig{gauss_vel}).  This can
be expressed by an integral
\begin{eqnarray}
f(v_{o},\sigma_{\phi} )  & = & \int_{-\infty}^{-v_o} \frac{e^{-v^2/(2\sigma_{\phi}^2)}}{\sqrt{2\pi} \sigma_{\phi}} dv \nonumber \\
& = & 0.5[1- Erf(\frac{1}{\sqrt{2}}\frac{v_o}{\sigma_{\phi}} )] \nonumber \\
& = & 1- I_{g}(\frac{v_o}{\sigma_{\phi}} ) \nonumber \\
& = & 1- I_{g}(\frac{1}{\zeta} ) 
\label{eq:f_v_sigma}
\end{eqnarray} 
where $I_{g}(x)= \int_{-\infty}^{x} \frac{e^{-t^2/(2\sigma^2)}}
{\sqrt{2\pi}\sigma}dt$ is a Gaussian integral.


$\lambda$ is related to $<j_z>$ by
\begin{equation}
\lambda     =  \frac{<j_z>}{\sqrt{2}V_v R_v }
\label{eq:lambda1}
\end{equation} 
If we further make the approximation that $<j_z / r> = <j_z> / <r>$,
then using \equ{lambda1} we can write
\begin{eqnarray}
v_o   &  = & <u_{\phi}>  \nonumber \\       
      &  = & <j_z/r>  \nonumber \\       
      &  \sim & <j_z> / <r>   \nonumber \\       
      &  \sim & \frac{ \sqrt{2} \lambda/(V_v R_v)}{ <r>} \nonumber \\       
      &  \sim & \frac{ \sqrt{2} \lambda V_{v}}{ <r/R_{v}>}       
\end{eqnarray} 
This can be used to express $f$ in terms of $\lambda$ as shown below
\begin{eqnarray}
f & = &1-I_{g}( \frac{\sqrt{2} \lambda}{ <r/R_{v}> (\sigma_{\phi}/V_{v})} )  \nonumber \\
 &=& 1-I_{g}( \frac{\sqrt{2} \lambda}{ k_{r} k_{\sigma}} )
\label{eq:f_kr}
\end{eqnarray}
where $k_{r}=<r/R_{v}>$ and $k_{\sigma}=(\sigma_{\phi}/V_{v})$ are
defined to be two quantities that are constant for a halo.  For an NFW
halo with concentration parameter $c$
\begin{eqnarray}
k_{r} & = & <r/R_{v}>  \nonumber \\
      & =  & \frac{ \int r \rho({\bf r})d^{3}{\bf r}}
{ \int \rho({\bf r})d^{3}{\bf r}}  \nonumber \\
      & =  & \frac{\pi}{4}\frac{[1+c-1/(1+c)-2\ln(1+c) ]}
{c[1/(1+c)-1+\ln(1+c)]}  \nonumber \\
      & =  & \frac{\pi}{4}f_{1}(c)  \nonumber \\
      & \sim  & 0.32 \ \ \ \ \ \mathrm{for} \ c=10.0 
\label{eq:k_r}
\end{eqnarray}

We define
\begin{eqnarray}
\lambda_{0}=<k_{r}> <k_{\sigma}>/\sqrt{2}
\label{eq:lambda_0}
\end{eqnarray}
then
\begin{eqnarray}
f & = & 1-I_{g}( \frac{ \lambda}{ \lambda_{0}} ) 
\label{eq:f_lambda}
\end{eqnarray}

\subsubsection{Anti-correlation of $f$ with $\lambda$}

\begin{deluxetable}{lllllll}
  \tablecaption{$k_r$ , $k_{\sigma}$ and $\lambda_0$ for Gas}
  \tablehead{ \colhead{ } & \multicolumn{2}{l}{$<k_r>$} &
    \multicolumn{2}{l}{$<k_{\sigma}>$}  &  \multicolumn{2}{l}{$\lambda_{0}$ } \\
    \cline{1-7}  \\
    \colhead{z} & \colhead{Sim} & \colhead{ENS} & \colhead{Sim} &
    \colhead{ENS} & \colhead{pred} & \colhead{Sim}} \startdata
  0    &  0.308 &  0.315 &  0.163 & -  &  0.036 &  0.036       \\
  1    &  0.343 &  0.349 &  0.181 & -  &  0.044 &  0.053       \\
  2    &  0.371 &  0.373 &  0.210 & -  &  0.055 &  0.063       \\
  3    &  0.380 &  0.390 &  0.217 & -  &  0.058 &  0.074       \\
  4    &  0.395 &  0.402 &  0.213 & -  &  0.059 &  0.074       \\
  \enddata
\label{tb:k_gas}
\end{deluxetable}
\begin{deluxetable}{lllllll}
  \tablecaption{$k_r$ , $k_{\sigma}$ and $\lambda_0$ for Dark Matter}
  \tablehead{ \colhead{ } & \multicolumn{2}{l}{$<k_r>$} &
    \multicolumn{2}{l}
    {$<k_{\sigma}>$}  &  \multicolumn{2}{l}{$\lambda_{0}$ } \\
    \cline{1-7}  \\
    \colhead{z} & \colhead{Sim} & \colhead{ENS} & \colhead{Sim} &
    \colhead{ENS} & \colhead{Pred} & \colhead{Sim}} \startdata
  0    & 0.308 &  0.315 &  0.632 &  0.654 &  0.138   &  0.145       \\
  1    & 0.350 &  0.349 &  0.613 &  0.605 &  0.152   &  0.170       \\
  2    & 0.382 &  0.373 &  0.595 &  0.578 &  0.161   &  0.178       \\
  3    & 0.398 &  0.390 &  0.590 &  0.561 &  0.166   &  0.177       \\
  4    & 0.409 &  0.402 &  0.573 &  0.551 &  0.166   &  0.189       \\
  \enddata \tablecomments{Sim: simulations, ENS: calculated
    theoretically by using \equ{k_r} and \equ{k_s}, $c$ used in
    calculations is estimated by algorithm given in
    \citet{2001ApJ...554..114E} $\lambda_0$ Pred: $\lambda_0$ as
    predicted by the toy model \equ{lambda_0}, the values of $<k_r>$
    and $<k_{\sigma}>$ used are the ones obtained from simulations,
    $\lambda_0$ Sim: $\lambda_0$ as obtained by fitting \equ{f_lambda}
    to the $f$ vs $\lambda$ data from simulations (\fig{gl-prop5}).}
  \label{tb:k_dm}
\end{deluxetable}

\equ{f_lambda} explains the anti-correlation of $f$ with $\lambda$ as
discussed in Section 3.2.3. Rightmost column in \tab{k_gas} and
\tab{k_dm} lists the values of parameter $\lambda_0$ obtained by
fitting this equation to the data from simulated halos (
\fig{gl-prop5}).  $\lambda_0$ as predicted by the toy model can be
calculated by using \equ{lambda_0}. For this first $k_{r}$ and
$k_{\sigma}$ are calculated for each of the halos in the simulation
then at any given redshift $<k_{r}>_{sim}$ and $<k_{\sigma}>_{sim}$
are calculated by taking the mean over all the halos at that redshift.
These values are listed in columns 2 and 4 of the aforementioned
tables.  $\lambda_0$ calculated from these values is listed in column
6. The values of $ \lambda_0$ predicted by the toy model have a small
offset but otherwise they are in agreement with those obtained from
simulations. The increase of $\lambda_0$ with redshift can be
understood in terms of variation of $<k_r>$ and $<k_{\sigma}>$ with
redshift.  $k_r$ as given by \equ{k_r} is a monotonically decreasing
function of $c$.  $c$ on the other hand has a dependence on mass,
redshift and cosmology given by
\begin{equation}
\label{eq:ENS}
c=c_{ENS}(\sigma_8,\Gamma,\Omega_{\Lambda},\Omega_{0},z,M_{f},M_0)
\end{equation}
which can be calculated by means of an algorithm given in
\citet{2001ApJ...554..114E} . For calculating $<c>$ at a given
redshift $z$ we put in \equ{ENS} $M_0=<M_v>=$the mean mass of a halo
at that redshift.  The predicted ENS values of $<k_r>$ are shown in
the table, they are in good agreement with those obtained from
simulations.  For the case here $<c> \sim 12.0/(1+z)$ (in agreement
with redshift dependence given in \citet{2001MNRAS.321..559B}). This
is the cause for increase of $<k_r>$ with redshift both for gas and
DM.  $k_{\sigma}$ for DM also depends on $c$ because for DM the one
dimensional velocity dispersion is given by $\sigma^2 = (1/3)V_{v}^2
f_c $, where $f_c \sim 2/3+(c/21.5)^{0.7}$ ( assuming that the
velocity distribution is isotropic and homogeneous, see
\citet{1998MNRAS.295..319M}).
\begin{eqnarray}
k_{\sigma}  &   =   & \frac{\sigma}{V_v} = \sqrt{f_c/3}  \sim 0.65 \ \ \ \ \ \ \ \ \mathrm{for}\ c=10.0  
\label{eq:k_s}
\end{eqnarray}
The slight decrease of $k_{\sigma}$ with redshifts (\tab{k_dm}) is
again due to increase of $<c>$ with redshifts.  For DM $k_r$ decreases
with $c$ while $k_{\sigma}$ increases this makes $\lambda_0$ nearly a
constant for all the halos at a given redshift and this is one of the
reason for the small scatter in $f$ vs $\lambda$ plots for DM (column
3 \fig{gl-prop5}).  For gas $\sigma_{\phi}$ is not related to $V_{v}$
and this results in a large scatter seen in its $f$ vs $\lambda$ plots
(column 3 \fig{gl-prop5}).  But $k_{\sigma}$ can be written in terms
of $F_{Tr}$ as $<k_{\sigma}> \sim \sqrt{F_{Tr}/3}$ and $<F_{Tr}>$
increases with redshift (Section 3.3 ,\fig{thermal1}), explaining the
increase of $k_{\sigma}$ with redshift shown in \tab{k_gas}.

It was found in B2001 that standard deviation $\sigma_j$ of angular
momentum for a subsample of $N$ particles for DM scales like

\begin{equation}
\frac{\sigma_J}{J}=\sqrt{1/N+1/(25\lambda^2N)} \simeq \frac{0.2}{\lambda\sqrt{N}}
\end{equation}
In the light of the toy model the error in $j$, for a sample of $N$
particles at some fiducial radius r, due to Poisson statistics scales
as $\sigma_j=\sigma_{\phi} r/\sqrt{N})$. So

\begin{eqnarray}
\frac{\sigma_j}{j}  & =  & \frac{\sigma r }{v_o r \sqrt{N}}  = \frac{\zeta}{\sqrt{N}}  
\end{eqnarray}
Comparing \equ{f_v_sigma} and \equ{f_lambda} $\zeta=\lambda_0/\lambda$
implying
\begin{eqnarray}
\frac{\sigma_j}{j}  & =  & \frac{\lambda_0}{\lambda\sqrt{N}}
\end{eqnarray}

$\lambda_0$ for DM as shown in table above is close to 0.2 which
explains the scaling relation observed by B2001.

\subsubsection{Distribution of $f$ and its change with redshift}
\begin{figure}
  \epsscale{1.1} \plotone{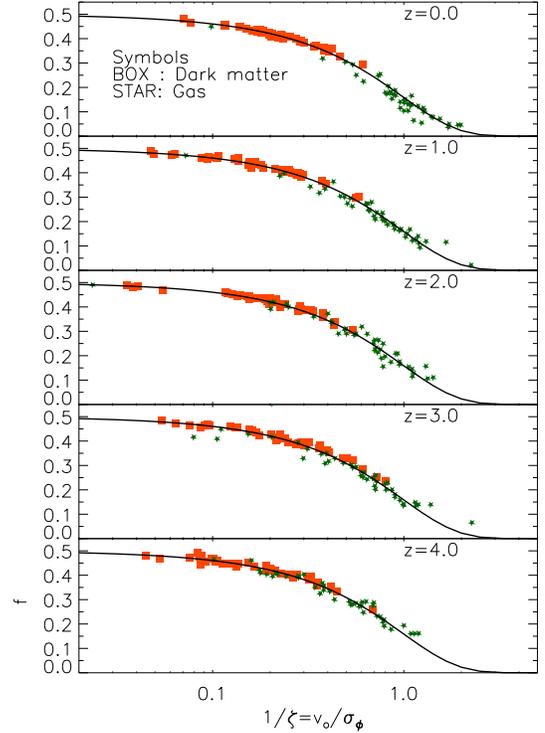}
  \caption{ Fraction of counter-rotating matter $f$ as a function of
    $1/\zeta=v_{o}/\sigma_{\phi}$ where $v_o$ is the mean tangential
    velocity and $\sigma_{\phi}$ its dispersion, the two parameters of
    the GSOV toy model described in Section 4. The solid line is the
    relationship $f=1-I_g(1/\zeta)$ as predicted by the model. Unlike
    the $f$ vs $\lambda$ plots where there was a large scatter of
    points about the theoretical curve for gas (right column
    \fig{gl-prop5}) and a free parameter $\lambda_0$ was required to
    fit the different cases, the data points for the $f$ vs $1/\zeta$
    plots for both gas and dark matter lie on the same predicted curve
    and they have a very small scatter.
\label{fig:gl-prop6}}
\end{figure}
The distribution of $\lambda$ at any particular redshift is a
lognormal distribution which can be specified by parameters
$\bar{\lambda}$ and $\sigma_{\lambda}$. By using \equ{f_lambda} we can
calculate the distribution of $f$ as shown below
\begin{eqnarray}
P(f)df  & = & \frac{P(\lambda)}{\frac{d}{d\lambda}f(\lambda,\lambda_0)} df
\label{eq:Pf}
\end{eqnarray}
This predicted formula describes the observed distribution in
simulations fairly well. The smooth curve in \fig{predict_fdist} is
the one predicted by the \equ{Pf}.  $P(\lambda)$ does not have a
strong redshift dependence but $f$ is related to $\lambda_{0}$
(\equ{f_lambda}) and $\lambda_{0}$ increases with redshift for gas.
This causes the profile of $P(f)$ for gas to shift towards smaller
values of $f$ at lower redshifts.  For dark matter $\lambda_{0}$ is
nearly constant so the profile does not show any significant change.
The fact that $<f>$ for gas decreases with decrease of redshift is not
only due to increasing thermalization of gas as shown in Section 3.3
but is also due to due to two other factors.  $<f>$ as given by
\equ{f_kr} depends on $<\lambda>$,$<k_{r}>$ and $<k_{\sigma}>$.
$<\lambda>$ increases, while $<k_{r}>$ and $<k_{\sigma}>$ decreases
with decrease in redshift. All of them act in the same direction to
decrease $<f>$.

\subsubsection{$f$ as a function of $v_o/\sigma_{\phi}$}
The relation between $f$ and $\lambda$ as given by \equ{f_lambda} is
an approximate one, the actual relation of $f$ is as given by
\equ{f_v_sigma}.  Moreover in the plots in \fig{gl-prop5} the gas
shows a large scatter. So we measure $v_{o}$ and $\sigma_{\phi}$ for
each halo and plot $f$ vs $v_{o}/\sigma$. These are shown as points in
\fig{gl-prop6}.  The theoretical prediction of the toy model as given
by \equ{f_v_sigma} is shown as a solid line on the same figure. The
large scatter which was seen in $f$ vs $\lambda$ plots of gas vanishes
and moreover there is no free parameter in this relationship.

\subsubsection{Calculating the AMDs by utilizing the toy model}
\begin{figure}
  \epsscale{1.1} \plotone{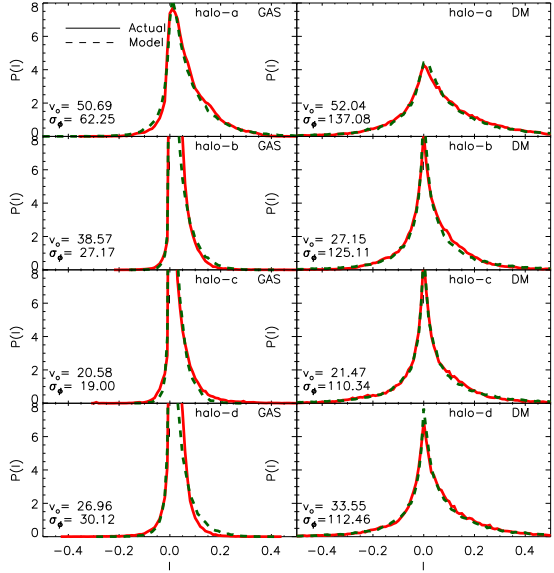}
  \caption{ Comparison of angular momentum distributions ($P(l)$ vs
    $l$ plots where $l=j/(\sqrt{2}R_vV_v)$) obtained from simulations
    with that of profiles obtained from the Gaussian smearing model
    (for four different halos same as the ones used in
    \fig{gauss_vel1}).  Solid line: profile from simulations Dashed
    line: profile generated by smearing model. The parameters $v_o$
    and $\sigma_{\phi}$ adopted for generating each of the model
    profiles is also indicated in each of the panels.
\label{fig:predict_amp3}}
\end{figure}

The model can also be used to calculate the AMDs.  Consider $m(r)dr$
to be the mass in a cylindrical shell of radius $r$ to $r+dr$. The
mean specific angular momentum at this radius is given by $v_{o}r$.
This is smeared by random motion with dispersion $\sigma_{\phi}$. So
the total mass with radius between $r$ to $r+dr$ and specific angular
momentum between $j$ to $j+dj$ is given by $
\frac{1}{\sqrt{2\pi}\sigma_{\phi}r}
e^{-(j-v_{o}r)^2/(2(\sigma_{\phi}r) ^{2})}dj m(r)dr$.  Integrating
this over $r$ we get the function $m(j)$ which is the mass of halo
with specific angular momentum between $j$ to $j+dj$.
\begin{equation}
  m(j) \ dj = \int^{R_{v}}_{0}  \frac{ 1}{ \sqrt{2\pi}\sigma_{\phi}r}  e^{-( j-v_{o} r )
    ^2/(2(\sigma_{\phi}r)^{2})} m(r) \ dr \  dj
\end{equation}
For a halo with a given $m(r)$ this integral can be calculated
numerically to give the distribution $m(j)$.  We instead follow an
alternative Monte-Carlo type approach due to its easier
implementation.  For a given halo once the model parameters $v_{o}$
and $\sigma$ are known we calculate the specific angular momentum of
each of the particles from equation $j=(v_{o}+v_{\sigma _{\phi}})r$
where $v_{\sigma_{\phi}}$ is drawn from a Gaussian distribution of
dispersion $\sigma$.  In \fig{predict_amp3} we have plotted the
angular momentum distributions obtained by the above procedure along
with the AMDs obtained from simulations. They are very similar.

\section{Generalized profiles based on Gamma Distribution}
\begin{figure}
  \epsscale{1.1} \plotone{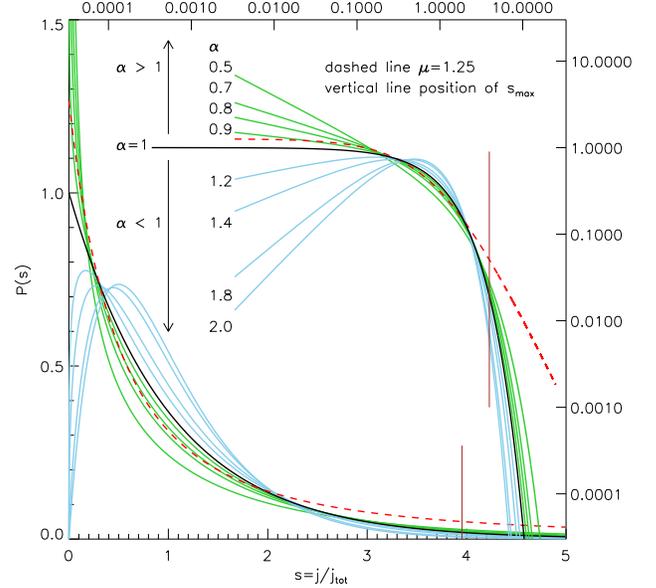}
  \caption{ The differential distribution of specific angular
    momentum for various values of $\alpha$. Both linear and log-log
    plots are shown. The dashed line corresponds to universal profile
    of \citet{2001ApJ...555..240B} with $\mu=1.25$ and the vertical
    line shows the position of $s_{max}$ for it. $\alpha$ is sensitive
    to the slope in the inner regions.  $\alpha=1$ marks the
    transition from a distribution that diverges as $s \rightarrow 0$
    ($\alpha < 1$) to the one that dips to zero ($\alpha > 1$).
\label{fig:gamma_profiles1_1}}
\end{figure}
\begin{figure}
  \epsscale{1.1} \plotone{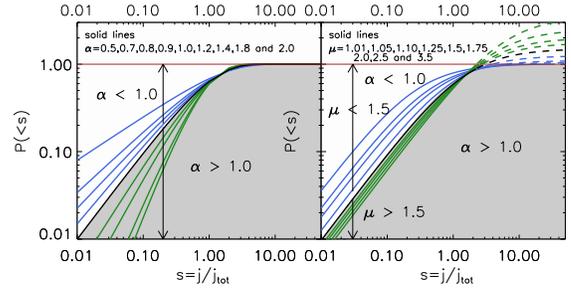}
  \caption{ The cumulative distribution $P(<s)$ of specific angular
    momentum $s=j/j_{tot}$ for various values of $\alpha$ (left) and
    $\mu$ (right). The shaded area separates the region with values of
    $\alpha>1$ form those with $\alpha < 1$ (unshaded area), in both
    the plots. The $\alpha=1$ curve roughly corresponds to curve with
    $\mu=1.5$.  The functional form of $\mu$ profiles can extend
    beyond $P(<s)=1$, this is shown as dashed lines. The profiles are
    truncated at $s=s_{max}$ the point where $P(<s)=1$.  For $\alpha$
    profiles $P(<s)$ is always less than 1 and it asymptotically
    approaches 1 for large values of $s$.  $\alpha$ and $\mu$ measure
    slightly different aspects of the shape. $\mu$ curves are a power
    law with slope of 1 for small $s$ and the value of $\mu$ is a
    measure of the point where the bend takes place. $\alpha$ on the
    other hand is sensitive to the slope of the curve for small $s$.
\label{fig:gamma_profiles1_2}}
\end{figure}
The various AMDs that have been analyzed in the previous sections can
be well described by an analytical function that depends on just one
parameter.  The functional form for the differential distribution is
based on the gamma distribution and reads

\begin{eqnarray}
  P(j) & = & \frac{1}{j_{d}^{\alpha}\Gamma(\alpha)}(j)^{\alpha-1}e^{-j/j_{d}} 
\end{eqnarray}
Since it is normalized it satisfies $\int_{0}^{\infty}P(j)dj=1$. Using
the fact that $\int_{0}^{\infty}jP(j)dj=
J/M=j_{tot}=\sqrt{2}R_{vir}V_{vir} \lambda $ we get

\begin{eqnarray}
  j_{d}  & = & \frac{j_{tot}}{\alpha}
\end{eqnarray}
This makes the distribution a one-parameter fit. We choose $\alpha$ as
the parameter as its effect is easy to understand on physical terms.
The differential distribution can be integrated to get the cumulative
distribution as shown below.

\begin{eqnarray}
  P(<j) & =& \frac{M(<j)}{M_v}  =  \int_{0}^{j}P(j)dj = \gamma(\alpha,j/j_{d})
\end{eqnarray}
where $\gamma$ is the Incomplete Gamma function. Writing in terms of
$s=j/j_{tot}$ and replacing $j_{d}$ we get

\begin{eqnarray}
  P(<s) & = & \gamma(\alpha,\alpha s)
\end{eqnarray}
\fig{gamma_profiles1_1} shows $P(s)$ vs $s$ plots for various values
of $\alpha$ in both linear and logarithmic plots.  For $s \ll
1/\alpha$ , $P(s)$ is a power law with slope $\alpha-1$. So for
$\alpha >1$ $P(s)$ goes to zero as $s \rightarrow 0$, and this gives
rise to a dip which is similar to the AMDs of observed dwarf galaxies
as shown in BBS01.  For $\alpha < 1$ the profiles diverge as $s
\rightarrow 0$ similar to AMDs seen in simulations.

\fig{gamma_profiles1_2} illustrates the effect of $\alpha$ on the
shape of $P(<s)$ vs $s$ plots, for reference we also illustrate the
effect of varying $\mu$ (B2001) on the shape of the profiles.

\subsection{What kind of AMD do spiral galaxies have?}
\begin{figure}
  \epsscale{1.1} \plotone{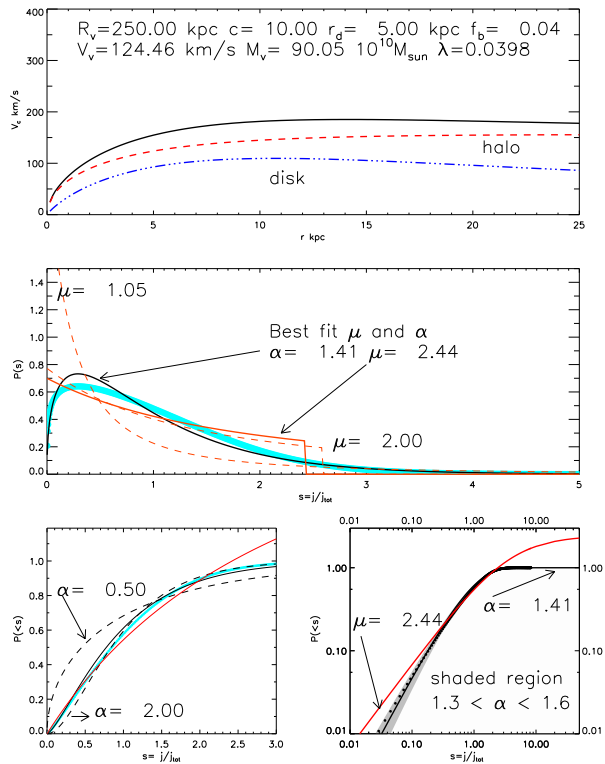}
  \caption{ The AMD of an exponential disk embedded in an NFW halo.
    Rotation curves are calculated by taking the adiabatic contraction
    of the halo and the flattened geometry of the disk into account.
    The thick light line (cyan) are the data points corresponding to
    AMD obtained from the model. A value of $\alpha=1.41$ seems to fit
    the model AMD (thin dark; black). $\mu$ profiles are not very
    suitable for describing the model profiles. The best fit $\mu$
    curve is shown as solid line (thin semi dark;red) and also for
    reference curves with $\mu=1.05$ and $\mu=2.0$ are also shown as
    dashed lines in the second panel. In lower-left panel we plot the
    cumulative distributions, the solid lines are the curves with best
    fit values of $\alpha$ and $\mu$. The dashed lines here are for
    curves with $\alpha=0.5$ and $\alpha=2.0$. In lower-right panel
    the shaded strip corresponds to the region where $1.3 < \alpha <
    1.6$, which is the expected range of values for model galaxies
    with $2<c<20$ ($r_d < 7 kpc$ and $R_v>200 kpc$). The dots are the
    AMD of the model.
\label{fig:nfw_disk1}}
\end{figure}
AMDs of observed dwarf galaxies as illustrated in BBS01 are found to
have a dip at low $s$ which suggests that they can be fit with a
profile having $\alpha > 1$. The simple case of an exponential disk
rotating in an isothermal halo ($\rho(r) \propto 1/r^2$) with flat
rotation curve can be calculated analytically and it gives a profile
with a value of $\alpha=2$. To investigate the AMDs of disk galaxies
in more detail and to find out the range of values of $\alpha$ that
they satisfy we create a model in which an exponential disk of mass
fraction $f_d$ and scale length $r_d$ is embedded in an NFW DM halo of
virial radius $R_v$ and concentration parameter $c$ ,similar to
analysis by \citet{1998MNRAS.295..319M}.  The disk mass fraction is
defined as $f_{d}=M_{disk}/(\Omega_{b} M_{v}/ \Omega_{m} )$.  For
calculating the rotation curve we take into account the adiabatic
contraction of the halo and also the flattened geometry of the disk.
So the input parameters for the model are $R_v,r_d,f_{d}$ and $c$.
\fig{nfw_disk1} shows a typical AMD obtained by this model.

The profile is very similar to that observed by BBS01.  It can also be
seen from the figure that the $\mu$ profiles are very different from
the AMD of model galaxies. For higher values of $\mu$ (about 2) they
can be an approximate fit but still they have a distinct core and tail
excess. They do not show the characteristic dip at low $j$ as shown by
models and also the tail has an abrupt truncation whereas the models
show a smooth extended distribution.  So a $\mu$ profile even with
higher values of $\mu$ does not describe the AMD of real galaxies. The
generalized profile described in Section 5 provides a good fit to
these model profiles.  For realistic values of the model parameters as
shown in \fig{nfw_disk1} we get $\alpha \sim 1.4$.  $\alpha$ is a
strong function of concentration parameter $c$ but only weakly related
to $r_d$ , $R_v$ and $f_d$ (keeping other parameters fixed). For lower
values of $f_d$ the fits get more and more accurate.  $\alpha$
increases from $1.30$ for $c=2$ to $1.6$ for $c=20$ (with $r_d<7 kpc$
and $R_v>200 kpc$).  We assume $1.3 < \alpha < 1.6$ as the typical
range of values expected for real galaxies.

\section{Techniques to measure the Angular Momentum Distribution of halos in simulations}
In this section we first describe two different methods to measure the
AMDs in halos followed by analysis of AMDs obtained by each of these
methods.  As described earlier the velocities given by simulations are
of two types.  The velocity of an SPH particle (gas) refers to the
streaming velocity $u$ while the velocity of a DM particle samples the
actual microscopic velocity $v=u+w$, where $w$ is the random motion.
One method to measure the angular momentum distribution is to take the
angular momentum of each particle and plot its distribution ({\bf
  particle method}). This inevitably gives a negative tail resulting
in a significant fraction of matter with negative AM.  For gas we find
it to be between $4\%$ to $48\%$ with a mean of about $15\%$ while for
DM it is between $30\%$ to $48\%$ with a mean of $41\%$. This not only
makes the interpretation of AMD obtained by the particle method
difficult but also the comparison between the AMDs of gas and DM. To
make both the components come to the same footing we can either {\bf
  broaden} the velocities of gas particles or {\bf smooth} the
velocities of both gas and DM particles over a fixed number of
neighbors. However, even smoothening the velocities over 1000
neighbors is not enough to suppress the negative angular momentum
tail.

B2001 obtain the AMD by dividing the halo into cells and then
calculating the angular momentum by averaging over all particles in
the cell ({\bf cell method}). These cells are then used to plot the
cumulative distribution. In the particle method described above it is
possible to derive both the differential and cumulative profiles while
in the cell method only latter is possible owing to the small number
of cells, typically around 60.  Furthermore, the cells are assumed to
have a spherically symmetric geometry.  Each cell covers a full $2\pi$
range in $\phi$ and they span the range of $(r/R_v , sin\theta)$ from
(0,0) to (1,1). The radial shells are spaced such that each of them
contains the same number of particles. The shells are then divided
into 3 azimuthal cells of equal volume between $sin \theta =0$ and
$1$. Positions with same $r sin\theta$ above and below the plane
belong to same cell. For our calculations we divide the halo into 60
cells with approximately constant number of particles in each of them.
We call this method { \bf symmetrical cell method}. The symmetrical
cell method is more effective in reducing the negative AM material
compared to particle method. However the resulting AMD might be biased
if the system is not axis symmetric. We therefore repeat the analysis
with a modified scheme which is free from any inherent symmetry. We
divide the halo into 4 radial shells with the $nth$ shell containing
$2n^2$ particle ($2,8,18$ and $32$). Each shell is divided into $n$
polar zones denoted by $l$.  Each $l$ region is divided into $2l+1$
azimuthal zones.  Finally regions above and below the $z$ plane belong
to different cells.  The radial and polar divisions are done such that
each cell contains a constant number of particles. We refer to this
method as { \bf normal cell method}.


\subsection{Analysis by cell method and Bullock profiles}
\begin{figure}
  \epsscale{1.15} \plotone{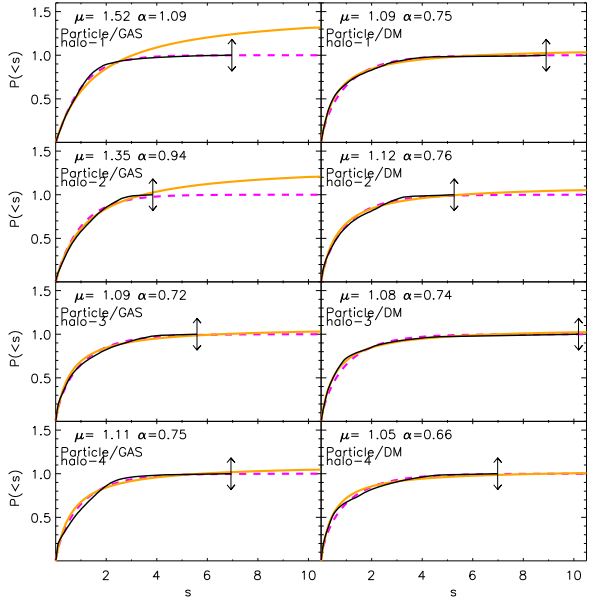}
  \caption{ The cumulative angular momentum distribution for 6
    different halos obtained by the particle method (dark solid line).  For each halo,
    distributions for both gas and dark matter are shown. The best fit
    $\alpha$ and $\mu$ values are also labeled . Light solid line (orange):
    best fit $\mu$ profile constant $\lambda$ fit.  Dashed line
    (magenta): best fit $\alpha$ profile.  Up down arrows mark the
    point with maximum value of $s$.The profiles have a smooth
    truncation the slope in outer parts gradually goes to zero.
    $s_{max}$ is higher than that obtained by cell method.
\label{fig:am_fits_lin_par}}
\end{figure}
\begin{figure}
  \epsscale{1.15} \plotone{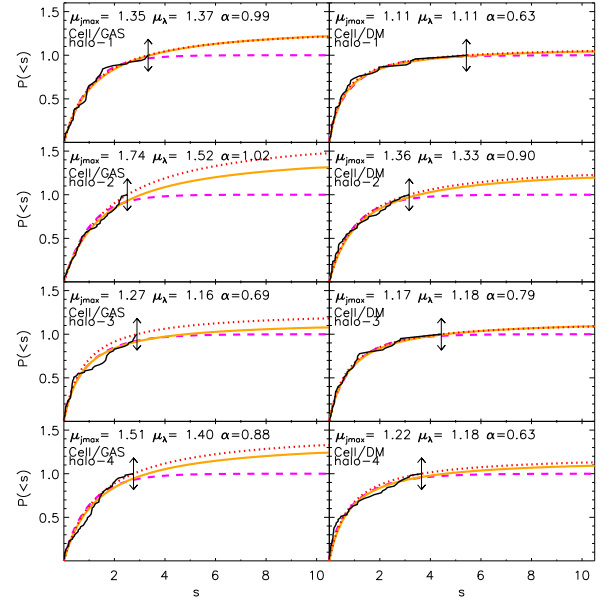}
  \caption{ The cumulative angular momentum distribution for 6
    different halos obtained by symmetrical cell method (dark solid line).  For each
    halo, distributions for both gas and dark matter are shown.  The
    best fit $\alpha$ and $\mu$ values are also labeled.  Light solid line
    (orange): best fit $\mu$ profile constant $\lambda$ fit.  Dotted
    line (red): best fit $\mu$ profile constant $j_{max}$ fit.  Dashed
    line (magenta): best fit $\alpha$ profile.  Up down arrows mark
    the point with maximum value of $s$.  The constant $j_{max}$ fits
    always pass through the point with maximum $s$ while the constant
    $\lambda$ and $\alpha$ profiles in general do not. The profiles
    tend to have an abrupt truncation, which means that at $s=s_{max}$
    the slope is significant.
\label{fig:am_fits_lin}}
 \end{figure}
\begin{figure}
  \epsscale{1.15} \plotone{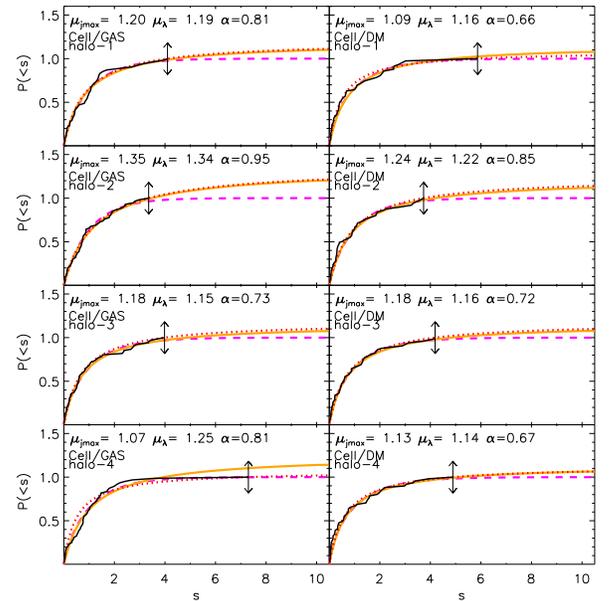}
  \caption{ The cumulative angular momentum distribution for 6
    different halos obtained by normal cell method free from any
    symmetry restrictions (dark solid line).  For each halo, distributions for both gas
    and dark matter are shown. The best fit $\alpha$ and $\mu$ values
    are also labeled. Light solid line (orange): best fit $\mu$ profile
    constant $\lambda$ fit.  Dotted line (red): best fit $\mu$ profile
    constant $j_{max}$ fit.  Dashed line (magenta): best fit $\alpha$
    profile.  Up down arrows mark the point with maximum value of $s$.
    The profiles end at higher values of $s$ and they also tend to
    have smoother truncation compared to profiles obtained by the
    symmetrical cell method. They are more similar to the profiles
    obtained by the particle method.
\label{fig:am_fits_lin_us}}
\end{figure}
\begin{figure*}
  \epsscale{1.1} \plotone{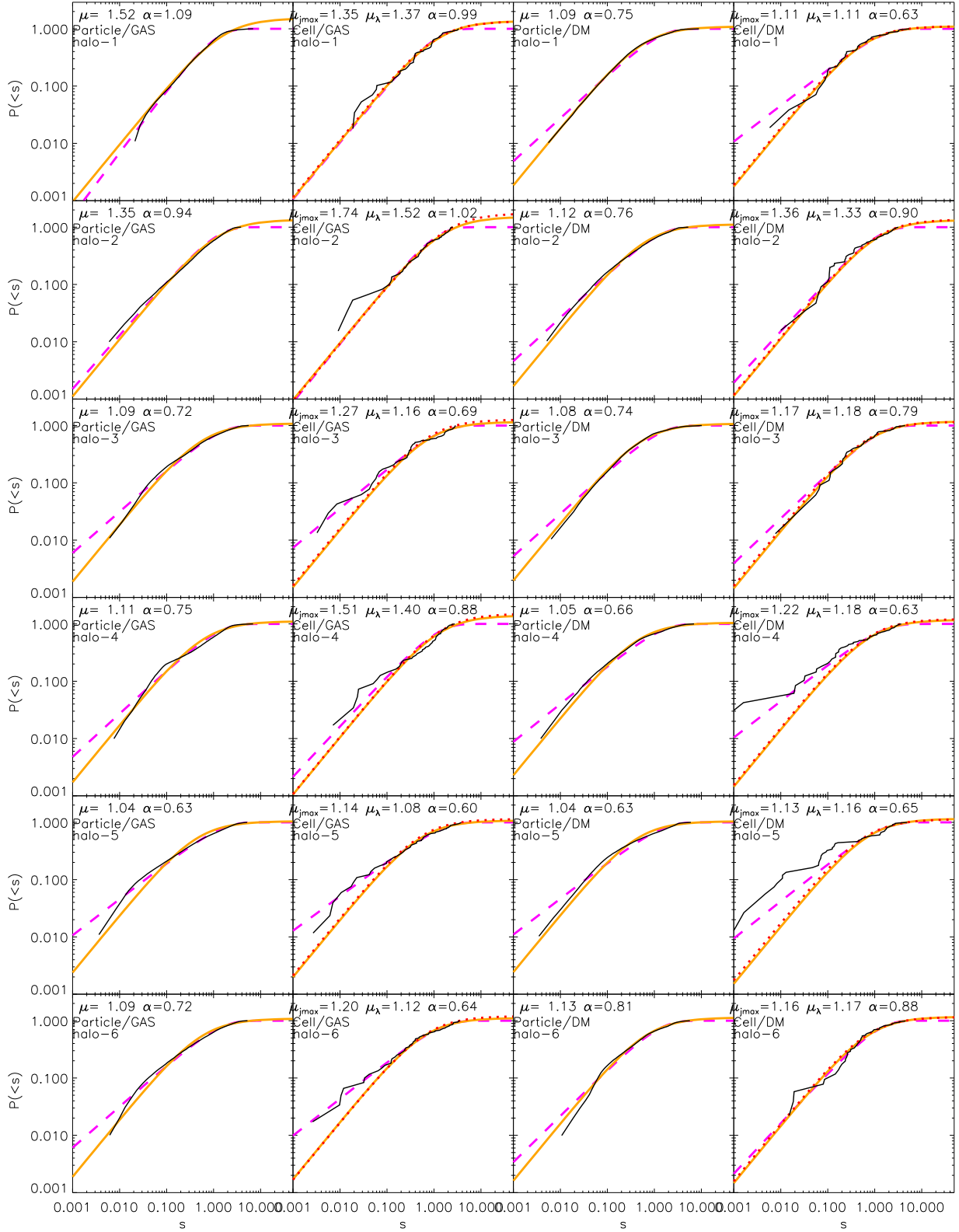}
\caption{ The cumulative angular momentum distribution for 6 different
  halos (dark solid line). Each row corresponds to a particular halo. Profiles of each
  halo both for gas and dark matter by particle and as well as
  symmetrical cell method are shown. Also are labeled the best fit
  $\alpha$ and $\mu$ values.  Light solid line (orange): best fit $\mu$
  profile constant $\lambda$ fit.  Dotted line (red): best fit $\mu$
  profile constant $j_{max}$ fit.  Dashed line (magenta): best fit
  $\alpha$ profile.  The subscript to $\mu$ denotes if it is a
  constant $\lambda$ or a constant $j_{max}$ fit.  For particle method
  the $\mu$ profiles are generated by constant $\lambda$ fits only.
\label{fig:am_fits}}
\end{figure*}

\begin{figure}
  \epsscale{1.1} \plotone{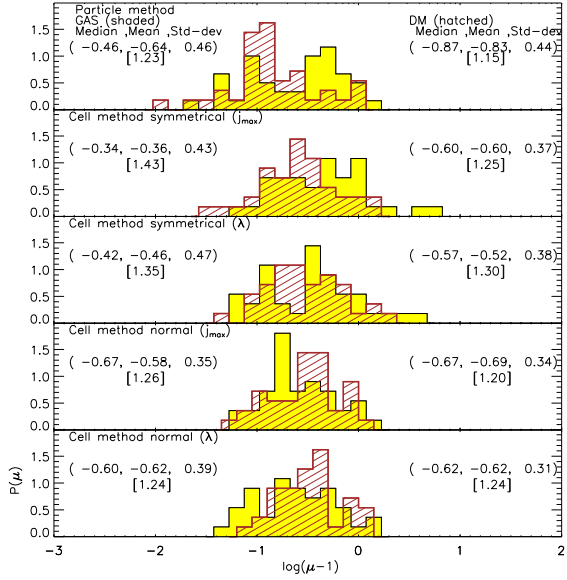}
  \caption{ Distribution of shape parameter $\mu$ for angular
    momentum distributions obtained by various different methods and
    fitting techniques. Median, mean and standard deviation of
    $log(\mu-1)$ is also indicated for each case, and below it
    corresponding $\mu$ value is shown.  For gas the constant
    $\lambda$ fits have lower mean and higher width compared to
    constant $j_{max}$ fits. On the other hand for DM the mean
    increases. Same trend is observed in analysis with normal cells.
    In comparison with symmetrical cells the normal cells give a lower
    mean and width for the $\mu$ distributions. The particle method
    gives even lower mean but the widths are maximum.
\label{fig:mu_dist}}
\end{figure}

In B2001 cumulative AMDs were calculated by the symmetrical cell
method for DM halos and a universal profile with shape parameter $\mu$
was shown to fit the data. The distribution of $\mu$ was found to
satisfy a log-normal distribution with $<log_{10}(\mu-1)>=-0.6$ and
$\sigma=0.4$. Our simulations reproduce these findings as shown in
\fig{mu_dist} (second row).  \citet[henceforth
CJ02]{2002MNRAS.336...55C} have also obtained similar results except
the fraction of cells with negative AM is higher than in B2001.  B2001
found that $5\%$ of halos have $f>0.1$ while CJ02 found it to be
$40\%$.  \citet[henceforth CJ03]{2003ApJ...597...35C} have also
analyzed the AMDs for both gas and DM components. They found that
$\mu$ is higher for gas than that for DM and also that $f$ is lower
for gas.  Results of our simulations agree quite well with the
findings of CJ02 and CJ03. We find that for DM $f>0.1$ for $33\%$ of
halo while for gas $f>0.1$ for $9\%$ of halos. $<f>$ for DM is $\sim
9\%$ while for gas it is $\sim 3\%$.  We also find $\mu$ to be higher
for gas compared to DM (second panel \fig{mu_dist}) by using the same
methods for analysis as done by above authors.

However, we observed that the $\mu$ values obtained from fitting
$M(<j)$ vs $j$ data are sensitive to the fitting procedure and error
bars used.  In fact two techniques can be used to fit $\mu$ profiles
each giving slightly different results.  The analytic function for the
$\mu$ profiles is given by
\begin{eqnarray}
\frac{M(<j)}{M_v} & = & P(<j)=\frac{\mu j}{j_{0}+j} 
\label{eq:mu_cumul1}
\end{eqnarray}
$j_{0}$ can be written in terms of $\lambda$ and $\mu$ as
\begin{eqnarray}
  j_{0} &=&\frac{j_{tot}}{b(\mu)}=\frac{\sqrt{2}V_{v}R_{v}\lambda}{b(\mu)}  
  \label{eq:j0lambda}
\end{eqnarray}
where $b(\mu)=-\mu \ ln(1-\mu^{-1})-1$. This makes the $\mu$ profiles
a one parameter fit for a halo with a given $\lambda$.
\begin{eqnarray}
  P(<j) & = & \frac{\mu j}{j_{0}(\mu, \lambda)+j} \\
  P(<s) & = & \frac{\mu s}{1/b(\mu)+s} 
\end{eqnarray}
We call this the { \bf constant ${\bf \lambda}$ fits}.  Alternatively
the $\mu$ profiles have an implicit maximum specific angular momentum
$j_{max}=j_{0}/(\mu-1)$. Writing in terms of $j/j_{max}$
\begin{eqnarray}
  P(<j) & = & \frac{\mu \ j/j_{max}}{(\mu-1)+j/j_{max}} 
\end{eqnarray}
For a given halo in simulations $j_{max}$ is the specific AM of the
cell having maximum specific AM. This is again a one parameter fit
which we call {\bf constant ${\bf j_{max}}$ fits}.

\begin{figure}
  \epsscale{1.1} \plotone{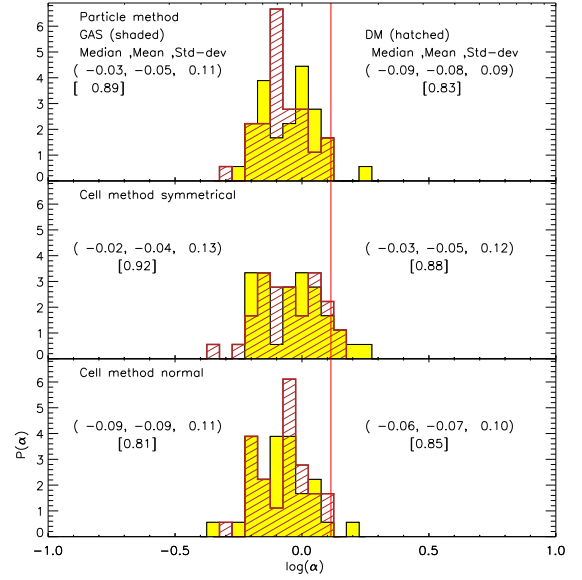}
  \caption{ Distribution of shape parameter $\alpha$ for angular
    momentum distributions obtained by the particle and by the cell
    method.  Median, mean and standard deviation of $log(\alpha)$ is
    also indicated for each case, and below it corresponding $\alpha$
    value is shown.  The vertical line corresponds to $\alpha=1.3$.
    The normal cell method gives lower values of mean and width for
    both gas and DM components compared to symmetrical cell. The same
    trend is observed for particle method except that the mean value
    for gas increases.
\label{fig:al_dist}}
\end{figure}
\begin{figure}
  \epsscale{1.1} \plotone{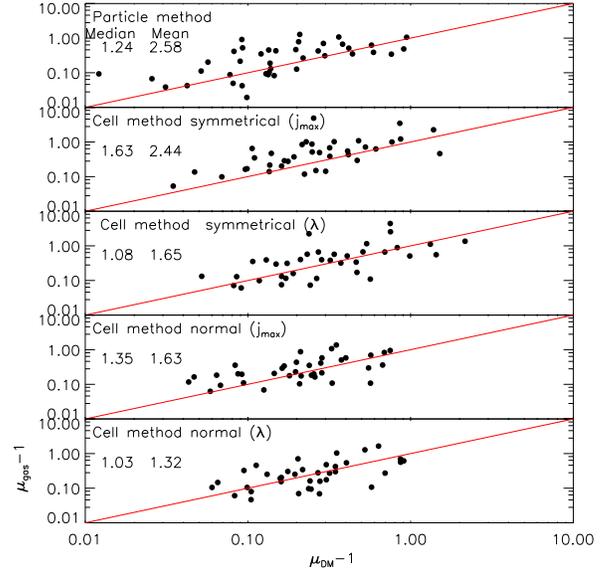}
  \caption{ $\mu_{Gas}-1$ vs $\mu_{DM}-1$ for various different
    techniques. Median and mean of $(\mu_{Gas}-1)/(\mu_{DM}-1)$ are
    also labeled on each of the plots.  The straight line shown in the
    plot corresponds to the relation $\mu_{Gas}=\mu_{DM}$.  Particle
    method and constant $j_{max}$ fits have $\mu_{Gas} > \mu_{DM}$ but
    constant $\lambda$ fits do not show any significant bias.
\label{fig:mu_prop}}
\end{figure}
\begin{figure}
  \epsscale{1.1} \plotone{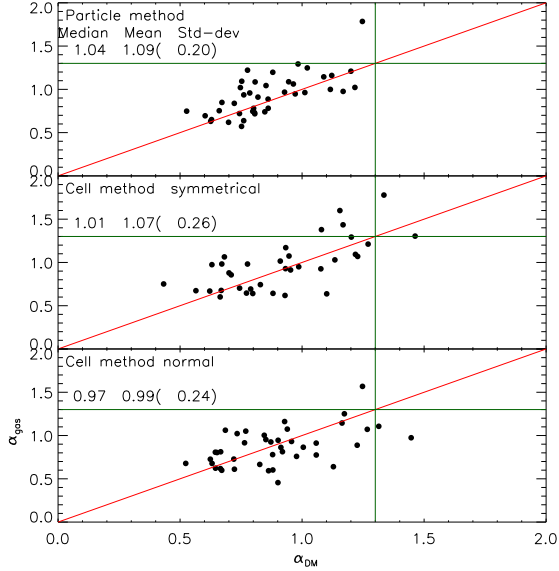}
  \caption{ $\alpha_{Gas}$ vs $\alpha_{DM}$ for various different
    techniques. Median ,mean and standard deviation of
    $(\alpha_{Gas}/\alpha_{DM})$ are also labeled on each of the
    plots.  The straight line shown in the plot corresponds to the
    relation $\alpha_{Gas}=\alpha_{DM}$.  In particle method we see
    $\alpha_{Gas} > \alpha_{DM}$ but cell method does not show such a
    bias. The two vertical and horizontal lines correspond to
    $\alpha=1.3$ and there are very few halos that lie above these
    lines (except for the middle panel which has a few).
\label{fig:al_prop}}
 \end{figure}

\begin{figure}
  \epsscale{1.1} \plotone{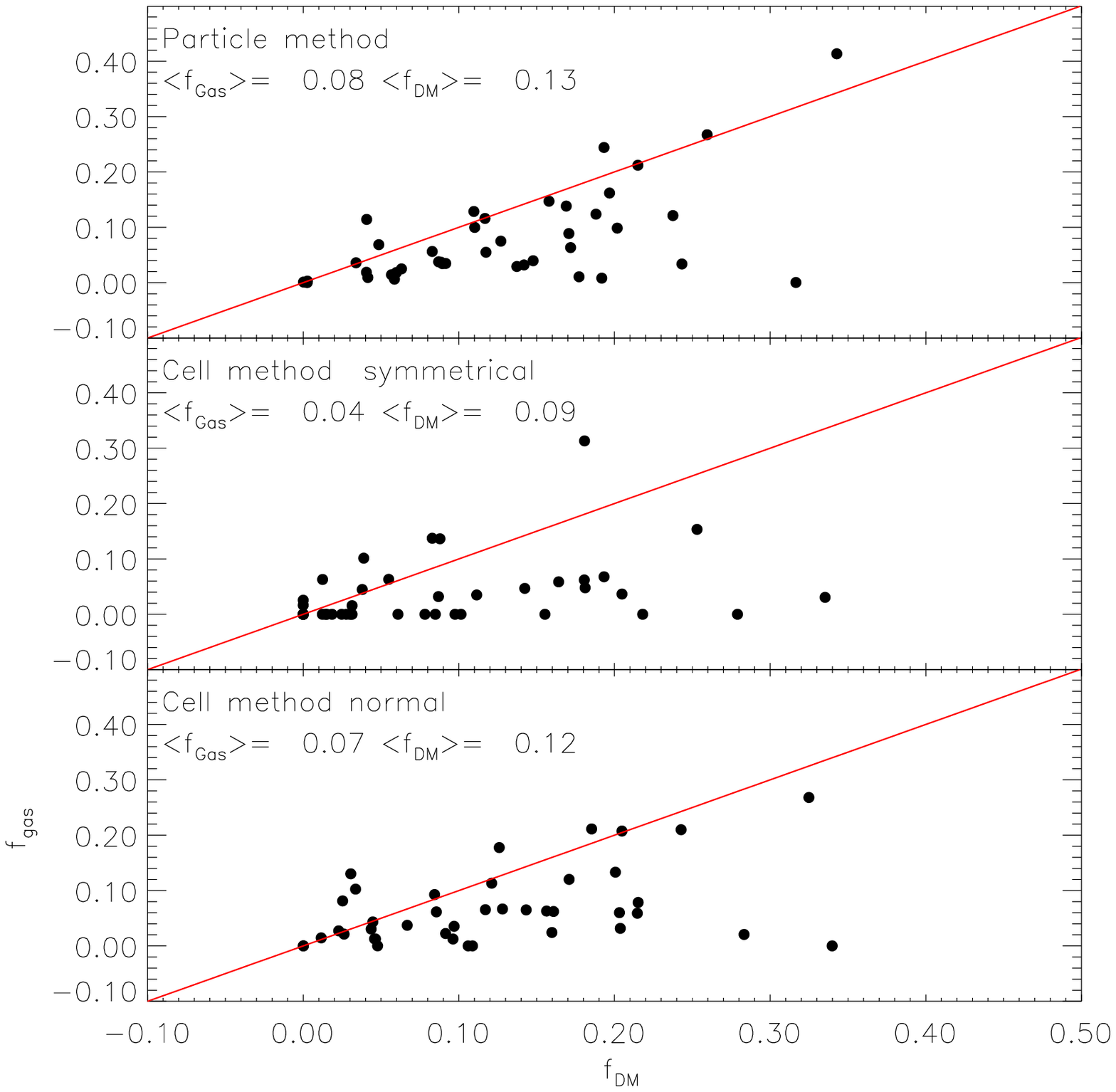}
  \caption{
    Comparison of parameters $f_{Gas}$ and $f_{DM}$ for various
    different techniques.  $<f_{gas}>$ and $<f_{DM}>$ are also shown.
    The distribution for normal cell method is very similar to that of
    particle method. In symmetrical cell method $f$ is in general much
    lower and for gas a significant number of the halos have $f=0$.
\label{fig:f_prop}}
\end{figure}

In B2001 it was shown that for DM the error bars on values of $j$ for
DM can be approximated by $\sigma_j=0.2/(\lambda \sqrt(N))$. The
velocity dispersions are in general not homogeneous throughout the
halo. CJ03 therefore base their analysis on error bars estimated by
$\sigma_j=v_c(r)r/\sqrt(N)$.  However, gas particles have a much
smaller velocity dispersion. We therefore use a scheme in which the
error are estimated by $\sigma_j=j_{std-dev}/\sqrt{N}$, where
$j_{std-dev}^2=<(j-<j>)^2>$ is the standard deviation calculated over
the particles in the cell.  For the cumulative distribution $P(<j)$ vs
$j$, $\sigma_j$ gives the error along j-axis. We estimate the error
along $P_{<}$ axis by $\sigma_{P_{<}}=(dP_{<}/dj)\sigma_j$.  To
calculate $dP_{<}/dj$ we first fit the data with a $\mu$ or $\alpha$
profile and then use this value of $\mu$ or $\alpha$ to calculate
$dP_{<}/dj$.  AMDs obtained by both particle and cell methods along
with corresponding $\mu$ and $\alpha$ fits are shown in
\fig{am_fits_lin_par}, \fig{am_fits_lin}, \fig{am_fits_lin_us} and
\fig{am_fits}.

We find that for constant $j_{max}$ fits $\mu_{gas} > \mu_{DM}$ in
agreement with CJ03 (\fig{mu_dist} and \fig{mu_prop} ,second row).
The distribution of $log(\mu-1)$ can be roughly fit by a Gaussian and
the fit parameters for DM are similar to those in B2001. For constant
$\lambda$ fits however, we find that $\mu_{gas}$ is nearly same as
$\mu_{DM}$. The reason for the discrepancy is that the constant
$j_{max}$ fits are constrained by construction to satisfy $P(<j)=1$ at
$j=j_{max}$, so the error in the value of $j_{max}$ is not taken into
account in the fitting procedure.  Moreover the fits are not
constrained to satisfy \equ{j0lambda}.  For constant $\lambda$ fits
and also the $\alpha$ fits this is not the case, the curves do not
necessarily truncate at $j=j_{max}$: they may extend to $j>j_{max}$ or
may already stop at $j<j_{max}$ (\fig{am_fits_lin}).  There is a
systematic trend such that for cases where $\mu$ is high (in constant
$j_{max}$ fits), the constant $\lambda$ fits truncate at $j>j_{max}$,
resulting in a lower value of $\mu$.  The DM does not have high values
of $\mu$ so it is relatively unaffected while gas has relatively high
values of $\mu$ and is significantly affected.  In other words, the
effect that $\mu_{gas} > \mu_{DM}$ is diminished to a large extent in
the constant $\lambda$ fits. The distributions of $\alpha_{gas}$ and
$\alpha_{DM}$ also show only a mild bias, (\fig{al_dist} second row ).
About $20\%$ of halos had $\alpha >1.3$ but out of these for $10\%$ of
halos the fits were poor so they were rejected.  After correction only
$10 \%$ of halos have $\alpha>1.3$ for gas, while for DM the
percentage is about $5\%$ (\fig{al_prop}).

Changing the geometry of the cells changes the distributions of $\mu$
and $\alpha$ slightly. The peak position and width of the
distributions are both reduced. A smaller percentage of halos have
$\alpha>1.3$.  We address the reason for these results in the next
section.

\subsection{Analysis by particle method}
\subsubsection{Is broadening technique suitable for comparing AMD of gas and DM? }
\label{sec:am_broad}
\begin{figure}
  \epsscale{1.1} \plotone{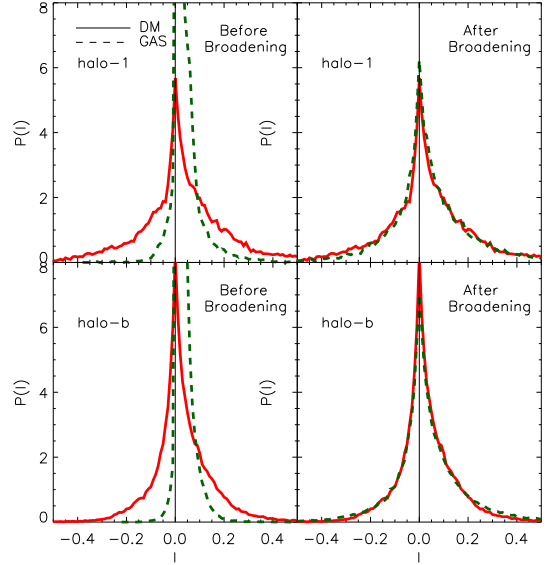}
  \caption{ Effect of broadening the velocities on the distribution
    of angular momentum for two different halos. After broadening the
    angular momentum distribution of gas is very similar to that of
    dark matter.
\label{fig:broad1}}
\end{figure}
\begin{figure}
  \epsscale{1.1} \plotone{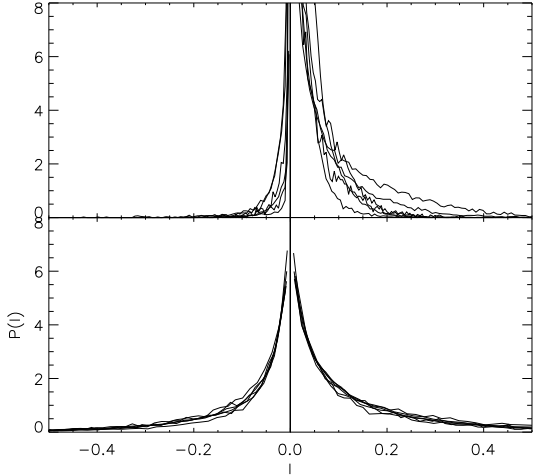}
  \caption{ Angular momentum distribution of gas of six different
    halos before broadening ({\bf top panel}) and after broadening
    their velocities with $\sigma=160.1 \kms $ (isothermal broadening)
    ({\bf lower panel}).  After broadening all the different
    distributions are similar so broadening is not a very reliable
    technique for making comparison between the distributions of gas
    and DM.
\label{fig:broad2}}
\end{figure}
It was shown in vB2002 that the AM profiles of gas and DM are
remarkably similar, if not identical, when the velocities of gas were
broadened by their microscopic thermal motion (\fig{broad1}). However,
if the shape of the profiles after broadening is determined primarily
by the value of $\sigma$ and is thus insensitive to the original
shape, then different profiles can be made to look similar by
broadening with same $\sigma$. We demonstrate this in \fig{broad2}
where we broaden the streaming velocities of 6 halos, each with a
unique AMD, with same velocity dispersion $\sigma=160.1 \kms $.
Although the initial profiles were quite different, the profiles after
broadening are very similar. We conclude that thermal broadening masks
out the uniqueness of un-broadened AMDs and hence is not a suitable
technique for making comparisons.

\subsubsection{Angular momentum distributions after smoothening}
\begin{figure}
  \epsscale{1.1} \plotone{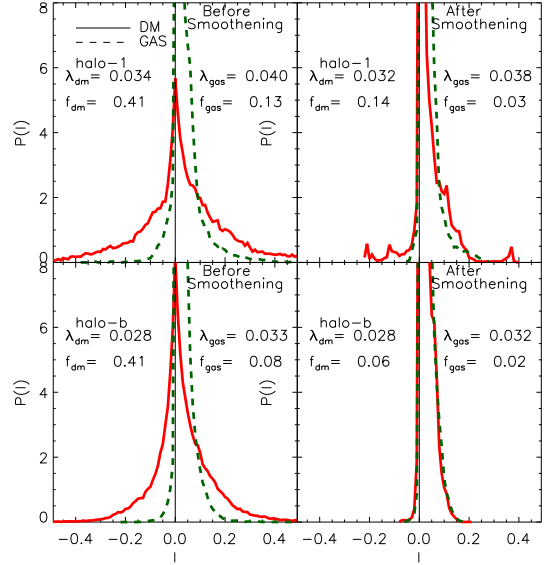}
  \caption{ Effect of smoothening on the angular momentum
    distributions of two different halos (same as in \fig{broad1}).
    The angular momentum of both gas and dark matter is smoothed by
    taking mean over 400 neighbors. For some halos after smoothening
    the gas and DM profiles are similar while for others they are
    different, one of the reasons for this is that the $\lambda$ and
    $f$ are not same for gas and DM.
\label{fig:smooth1}}
\end{figure}
\begin{figure}
  \epsscale{1.1} \plotone{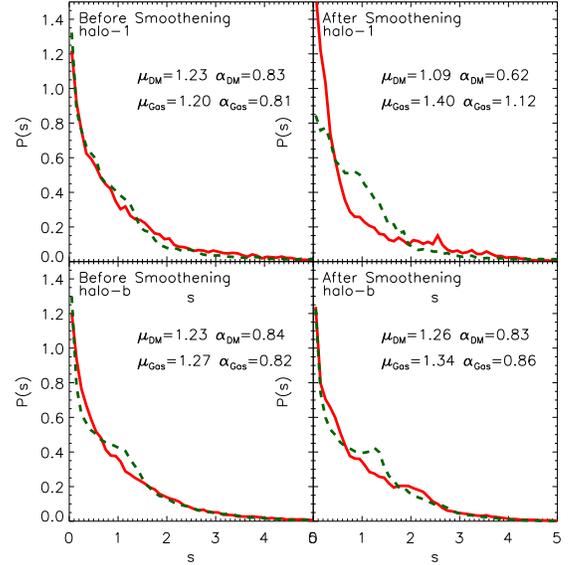}
  \caption{ Effect of smoothening on the $P(s)$ vs $s$ plots of two
    different halos (same as in \fig{broad1}).  The angular momentum
    of both gas and dark matter is smoothed by taking mean over 400
    neighbors. For the halo in the lower panel, the gas and DM
    profiles after smoothening are similar but for the other halo they
    are different.
\label{fig:smooth2}}
\end{figure}
Rather than broadening we consider smoothening of the DM velocities as
the superior procedure, i.e.\ we extract streaming velocity $u$ from
the given velocity $v$ and then compare their AMD with that of gas.
Increasing the numbers of neighbors $N_{ngb}$, over which the
smoothing is applied, will decrease $f$ but then the region over which
averaging is done eventually becomes too large and any significant
information is lost.  Furthermore, particles close to the rotation
axis suffer an apparent loss of AM due to a geometrical effect:
$<v_x>$ and $<v_y>$ in the inner region are zero for a symmetric
rotating system aligned along $z$ axis. Alternatively we may smooth
the AM instead ( the halo should be centered before smoothening). In
this case we encounter the following problem: Particles close to the
axis and in a conical region around it get enhanced in AM while those
along the equator suffer a loss in AM due to the existence of a strong
density gradient.  A spherical volume around such a typical particle
has more neighbors towards the symmetry axis with lower AM than away
from it, so the smoothed AM which has a radial dependence given by
$j(r)=v_o r$, is lower close to the axis. This effect was not
prominent when the smoothening was done on velocities because the
velocity does not exhibit a strong radial dependence ($v(r) \sim
v_o$). This suggests that a better method would be to smooth the
velocities in cylindrical co-ordinates. For our analysis here we use
the technique of smoothening the angular momentum only and we choose
the number of neighbors to be 400. Instead of an SPH type kernel we
use a simple step function, which is equivalent to taking the mean
over the neighbors. This makes it easier to make comparisons with the
cell method and is also more effective in reducing $f$.

In(\fig{smooth1}) the effect of smoothening on the distributions of
angular momentum is shown. For some halos the profiles of gas and DM
are similar after smoothening and for some they are different. This
kind of comparison is not very useful because the shape of $P(l)$ vs
$l$ plots depends on $\lambda$ and $f$. So a good agreement may merely
reflect that the $\lambda$ and $f$ are similar for gas and DM.
Moreover $\lambda$ and $f$ are in general not same for gas and DM,
this makes the interpretations of these plots even more difficult.

So to make a comparison first we need to take out the dependence on
$\lambda$ and $f$ which can be done by plotting $P(s)$ vs $s$ for the
positive tails of gas and DM and then comparing the best fit values of
$\mu$ or $\alpha$. This is shown in \fig{smooth2}.

\subsubsection{Analysis of smoothened profiles}
\begin{figure}
  \epsscale{1.1} \plotone{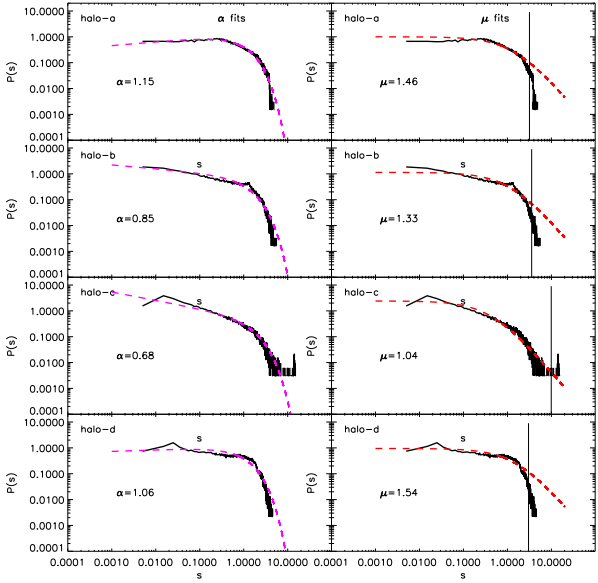}
  \caption{ A comparison of $\mu$ and $\alpha$ fits applied to
    angular momentum distributions of the gas obtained by the particle
    method. The vertical line in plots on the right indicate the
    truncation point $s_{max}$ of $\mu$ profiles. The solid lines
    are the data obtained from simulation and the dashed lines
    are the best fit $\alpha$ (left) and $\mu$ (right) profiles. In
    outer parts the actual profiles are steeper than $1/s^2$ (the
    expected asymptotic form of $\mu$ profiles).
    \label{fig:smooth_effect3}}
 \end{figure}
 We bin the smoothed profiles such that each bin contains
 $N_{halo}/100$ particles. For calculating the error bars we use the
 same technique as described earlier in the case of the cell method,
 with the exception that we assume $\sigma_j/j = <\zeta>/\sqrt{N}$.
 For DM $<\zeta>=5$ and with $N=400$ this gives a value of 0.25. For
 gas $<\zeta>=1$ and this gives a value of 0.04.  The smoothed
 profiles can be fit by both Bullock $\mu$ profiles (only constant
 $\lambda$ fits are used) and the generalized $\alpha$ profiles
 \fig{smooth2}.  $P(s)$ vs $s$ differential plots in
 \fig{smooth_effect3} show that that the Bullock profiles are
 shallower for large $s$, i.e. the actual AMDs are steeper than
 $1/s^{2}$ in outer parts.  The particle profiles do not truncate
 abruptly like in the cell method, the slope gradually goes to zero.
 The $\alpha$ profiles provide better fit to the particle profiles in
 outer regions.

 The $\mu$ and $\alpha$ values obtained by particle method are also
 larger for gas as compared to that of DM (\fig{mu_prop} and
 \fig{al_prop}).  $log(\mu-1)$ and $log(\alpha)$ distributions are
 roughly Gaussian. The mean, median and standard deviation are shown
 in \fig{mu_dist} and \fig{al_dist}.  $\alpha>1.3$ for about $ 2\%$ of
 gas halos while none of the DM halos have $\alpha>1.3$. Even after
 smoothening with same number of neighbors in particle method $f$ on
 average is greater for DM ($13\%$) than for gas ($8\%$).  Similar
 effect is also seen in cell method where the cell size used is same
 for both DM and gas \fig{f_prop}.

\subsubsection{Effect of spherical and symmetric cell geometry}
A comparison of results of particle method with that of symmetrical
cell method show that $\mu$ (constant $\lambda$ fits) and $\alpha$ are
higher for symmetrical cell method. $\mu$ specifically is higher
because $s_{max}$ to which it is directly related is about $1.5-2.0$
times higher in particle method than in symmetrical cell method
(\fig{am_fits_lin_par},\fig{am_fits_lin}) although the number of
neighbors in particle method is approximately the same as the number
of particles in each cell. $f$ is also significantly lower in
symmetrical cell method than in particle method (\fig{f_prop}).  Since
smoothening is a local effect whereas cell averaging is not due to its
spherical and symmetric construction. So a blob of negative AM or
higher AM material will retain its character when smoothed but will be
averaged out in cell method over regions that are not necessarily
local. Both methods will deliver identical results only in the case
where the system is symmetrical. Any deviation from symmetry, in the
symmetrical cell method has the effect of lowering the low AM material
by mixing it with high AM one and vice versa. So both $f$ and
$s_{max}$ are lower in symmetrical cell method.  $\alpha$ does not
depend upon $s_{max}$ but is only affected by the lowering of low AM
material and this effect is not very strong so it shows only a slight
change.

These arguments also imply that the results of normal cell method
should be very similar to particle method.  This is indeed the case as
can be seen in \fig{f_prop} where distribution of $f$ for normal cell
method is very similar to particle method.  The distribution of
$\alpha$ and $\mu$ parameter for particle method is also closer to
normal cell method than the symmetrical cell method as shown in
\fig{al_prop}. In spite of the similarities there are subtle
differences specially for gas whose $<log(\alpha)>$ is much lower for
the case of normal cell method.  This discrepancy may be because the
number of particles in the normal cell is not fixed for all halos like
in particle method for which the number is 400.  Compared to normal
cell method the particle method has slightly lower values of $\mu$
because their $s_{max}$ is slightly higher and this is because AM is a
monotonically increasing function of radius and for a cell of finite
radial thickness the average AM of the cell will always be less than
the maximum AM of particles in it.

\subsubsection{Measuring the spatial asymmetry of angular momentum
  distribution within a halo}

The above discussion suggests that both gas and DM should have
significant asymmetry. We verify this as follows: We measure the
asymmetry by dividing the halo into cells as described earlier but
this time each radial shell is also divided into 6 azimuthal zones and
3 $\theta$ zones.  Then we measure the $z$ component of angular
momentum vector $ j(r)$ and $j(r')$ for a pair of cells with opposite
parity situated at ${\bf r}$ and ${\bf r'}$ where ${\bf r'=-r}$.
Symmetry of $j$ distribution within a halo is given by
\begin{eqnarray}
  S_{j} & = & < \frac{ |j(r)+j(r')|^2-|j(r)-j(r')|^2 }{|j(r)+j(r')|^2+|j(r)-j(r')|^2} >_ \mathrm{all \ cells}   \nonumber \\
  & = & < \frac{2j(r)j(r')}{j^2(r)+j^{2}(r')} >_\mathrm{all \ cells}  
\end{eqnarray}
$S_j$ can vary from -1 (perfectly anti-symmetric system) to 1
(perfectly symmetric system).  An asymmetry as high as $j(r)=2j(r')$
corresponds to $S_j$ of only 0.8.  $<S_j>$ ,the mean over all halos,
is found to be 0.78 for DM and 0.83 for gas.  So in halos both gas and
DM have significant asymmetry as expected. DM in fact is more
asymmetrically distributed than gas.

\subsection{Comparison of quality of $\alpha$ and $\mu$ fits}
It might sound strange that two different functional forms are being
used to describe the same data. But it is not very surprising
considering the fact that the profiles cannot be perfectly described
by a functional form, they have small deviations and this gives enough
room to both $\mu$ or $\alpha$ profiles (which after all are not very
different) to be used to fit them with equally good fits.  The
profiles from the cell method (symmetrical) tend to have an abrupt
truncation at $j=j_{max}$ for some cases ( probably due to the
symmetry effects discussed in previous section and also because the
number of data points used to sample the profile, which is the total
number of cells is small, typically around $60$ ). So $\mu$ profiles
which by design have an abrupt truncation at $j=j_{max}$ perform
better in the outer regions for these cases, but for the same reason
they are not able to fit the particle profiles in the outer regions
which have a smooth extended tail. The $\alpha$ profiles fare better
for these. So if we neglect the outer parts $P(<s) > 0.95$ then both
the profiles provide a satisfactory fit to the distributions from both
methods.  To get an estimate of the quality of the fits, we measure
$\chi^2$ for data points with $P(<s) \le 0.95$ for both profiles.
$\chi^2$ is defined as $\chi^2= \sum_{i=1}^{N}
((y_i^{data}-y_i^{model})/\sigma_y)^2$ for each of the fits.  To make
a comparison we calculate the median of $(\chi_{\alpha}^2 -
\chi_{\mu}^2)/(\chi_{\alpha}\chi_{\mu})$ and median of
$\chi_{\alpha}^2/\chi_{\mu}^2 $ over all the halos.  If the $\chi^2$
for both of them are equal then the former quantity is close to zero.
A negative value of -1 implies $\chi_{\mu}^2=2.5\chi^{2}_{\alpha}$. If
the median of $\chi_{\alpha}^2/\chi_{\mu}^2 $ is close to 1 then both
fits are good for equal percentage of halos.  If more than $50\%$ of
halos have $\chi_{\alpha} < \chi_{\mu}$ then the above quantity will
be less than 1 and vice versa.  In \tab{chi} and \tab{corr} the values
of these quantities are listed for various methods.  Form the table we
can see that both the profiles are equally good at describing the AMDs
obtained by symmetrical cell and particle methods (except for the
outer $5\%$ ).

\begin{deluxetable}{lllll}
  \tablecaption{ Comparison of $\chi^2$ for $\mu$ and $\alpha$ fits}
  \tablehead { \colhead{ } & \multicolumn{2}{c}{Cell method
      \tablenotemark{a}}&
    \multicolumn{2}{c}{ Particle method} \\
    \cline{2-5} \\
    \colhead{ } & \colhead{DM} & \colhead{GAS} & \colhead{DM} &
    \colhead{GAS} } \startdata
  median($\chi_{\alpha}^2/\chi_{\mu}^2$)  &                0.78 &    1.31 & 1.47 & 0.92 \\
  median($\frac{\chi_{\alpha}^2-\chi_{\mu}^2}{\chi_{\mu}\chi_{\alpha}}$) &      -0.25 &    0.27 & 0.39 & -0.08 \\
  \enddata \tablenotetext{a}{Comparison shown for symmetrical cell
    method only}
  \label{tb:chi}
\end{deluxetable}

\begin{deluxetable}{lllllllll}    
  \tablecaption { { \scriptsize Pearson correlation coefficient of
      $\alpha$ with other halo parameters} } \tablehead { \colhead{} &
    \colhead{$M_v$} & \colhead{$ \lambda$} & \colhead{$c$} &
    \colhead{$S_j$} & \colhead{$\alpha_{cell}^{sym}$} &
    \colhead{$\alpha_{cell}^{norm}$} & \colhead{$\alpha_{DM}^{par}$} }
  \startdata
  $\alpha_{GAS}^{par}$  & -0.12  &  0.27       & -0.14  & -0.04 & 0.51                 & 0.86                   &  0.69           \\
  $\alpha_{DM}^{par} $   & -0.16  &  0.53       & -0.25  &  0.27 & 0.76                 & 0.80                   &  1.00           \\
  \enddata \tablecomments { The superscript on $\alpha$ denotes the
    method employed to calculate the AMD e.g particle method ,
    symmetrical cell method or normal cell method (free from symmetry
    restriction).  }
  \label{tb:corr}
\end{deluxetable}

\section{CONCLUSION \&  DISCUSSIONS}

We have presented here results from the non-radiative hydrodynamical
simulation of 41 high resolution halos whose masses were selected to
span the range from dwarf to bright galaxies. Our investigation mainly
focused on the angular momentum properties of the halos and whether
there are systematic differences between gas and dark matter. Our
findings can be summarized as follows:

\begin{enumerate}
\item We investigated some of the global angular momentum properties
  like spin parameter $\lambda$, fraction of negative angular momentum
  $f$ and misalignment angle $\theta$.  We find the spin parameter of
  gas to be on average larger than that of DM and this effect is
  systematically more pronounced at lower redshifts.  At $z=0$
  $\lambda_{gas}/\lambda_{DM} \sim 1.4$, which is in agreement with
  the result reported by CJ02, but is not in agreement with vB2002 who
  do not find any such bias. The mean of the misalignment angle
  $\theta$ is $20^{\circ}$ which is again in agreement with CJ02 who
  get a value $23.5^{\circ}$ but is less than the value of
  $36.2^{\circ}$ obtained by vB2002.  Both these discrepancies could
  be due to the inclusion of a large number of low resolution halos in
  the analysis of vB2002.  The counter-rotating fraction $f$ is
  anti-correlated with $\lambda$ and for gas $f$ decreases with
  decrease of redshift, an effect that can be explained by the
  increasing level of thermalization at lower redshifts.  Other than
  this there is little evolution of other properties with redshift.
  
\item We find that the fraction of material with negative angular
  momentum can be described by the equation
  $f=1-I_g(\lambda/\lambda_0)$, with $I(x)$ being a Gaussian integral.
  To understand these effects we developed a toy model, where we
  introduce an ordered velocity $v_o$ which is smeared by means of
  Gaussian random motion with dispersion $\sigma$. This model
  reproduces the $f=f(\lambda)$ correlation and suggests an actual
  relation of $f=1-I_g(1/\zeta)$ where $\zeta=\sigma/v_o$. This
  relation is demonstrated to be in excellent agreement with the
  results from simulations. We also see that the gas gets more and
  more thermalized at lower redshift resulting in $\sigma/V_v$ to be
  smaller at lower redshifts, whereas $\lambda$ and concentration $c$
  increases at lower redshifts. All these effects are contributing to
  the decrease of the amount of gas with negative angular momentum.
  The model also reproduces well the shape of AMD observed in
  simulations.
  
\item We study the distribution of angular momentum in detail, and
  compare and contrast various different techniques used to derive
  angular momentum distributions. We first use the cell method as
  proposed by B2001 and reproduce the result of CJ03 that $\mu$ for
  gas is greater than that of DM.  We find that the results are
  sensitive to the very details of the method employed for fitting.
  In particular the effect of gas having higher $\mu$ is diminished to
  a large extent if fits are performed assuming a constant $\lambda$
  rather than a constant $j_{\rm max}$.  According to CJ03 $\mu$ for
  gas is comfortably in the range required by observation of disk
  galaxies which is ($\mu>1.75$).  We also find that about $30\%$ of
  halos have $\mu >1.75$.
  
\item By comparing the AMD found in the simulations with those of
  exponential disks we conclude that merely having $\mu >1.75$ is not
  a sufficient condition to match the angular momentum profiles of
  observed disk galaxies.  We find that a generalized profile, based
  on gamma distribution, with a single parameter $\alpha$ can be used
  to fit the AMD of model galaxies (exponential disks embedded in NFW
  halos) as well as AMD of gas and DM in simulations. $\alpha > 1.3$
  seems to resemble the profiles of dwarf galaxies shown in BBS01, a
  condition that is only obeyed by a small minority of halos (less
  than $10 \%$).  For fits based on the (symmetric) cell method only
  $10 \%$ of the halos have gas with $\alpha >1.3$ (after rejecting
  halos with bad fits).  For particle method and normal cell method
  the percentage is even lower, about $2\%$ of halos have
  $\alpha>1.3$. We find the particle method to be more robust, with
  data that is less noisy and also free from any artifical non local
  averaging or any symmetry assumptions. The profile in particle
  method do not have abrupt truncation like in cell method
  (symmetrical) and are more extended. This may have important
  implications for the truncation radius and extent of gas in real
  disks: In semi analytical models the distribution and extent of cold
  gas depends upon the AMD used for the models.  In
  \citet{2001MNRAS.327.1334V}(Fig-9 there) the models have an AMD with
  $\mu$ between 1.6 and 1.9, and the disks are predicted to have a
  sharp truncation ($R_{gas}/R_{HI} \sim 1$) of cold gas which is in
  disagreement with observed distribution of HI in galaxies
  ($R_{gas}/R_{HI} > 1.5$).  Here $R_{gas}$ is the maximum extent of
  gas with non zero surface density and $R_{HI}$ is the radius with
  surface density of $1 \Msun pc^{-2}$.
  
\item For the particle method, there is a significant fraction of
  counter-rotating matter which has been excluded in calculating
  $\alpha$. The final AMD (and therefore the predicted structure of
  the model galaxy) will depend strongly upon how this material
  eventually gets mixed up with the remaining portion of the halo,
  during the assembly of the galaxy.  We saw that the counter-rotating
  matter undergoes extensive mixing during the history of its
  formation in the hierarchical framework. But the situation we have
  here is slightly different as it concerns the fate of the
  counter-rotating gas in an isolated halo when subjected to collapse.
  In the absence of any information it is best not to assume any
  preferential mixing with low, or high AM material or also equal AM
  material in which case it might form a non rotating bulge as
  prescribed in vB2002. The most plausible prescription seems to be
  that of random mixing: During collapse the counter-rotating matter
  will follow a slightly different trajectory than normal matter
  around it and in the process it will get mixed by shocks with matter
  of various different AM along its path. This process does not seem
  to have any preference.  If we assume that the process of mixing
  will not change the distribution of $\alpha$ significantly, we can
  conclude that less than $10 \%$ of halos have $\alpha>1.3$. In other
  words only the absolute minority of halos have AM distributions that
  resembles that of an observed (dwarf or LSB) disk galaxy.  In the
  absence of any significant correlation of $\alpha$ with mass $M_v$
  or for that matter with any other halo parameter, this leads us to
  the general conclusion that a typical halo in $\Lambda CDM$
  simulation has far too much low angular momentum material to account
  for majority galaxies featuring a dominated disk component.
  
\item If the angular momentum of each element is conserved then low
  angular momentum material has to be preferentially discarded during
  the process of galaxy formation, this might be effected by means of
  supernova feedback during star formation that drives out gas from
  small halos that come randomly from all direction and contribute
  mainly to the low angular momentum material as shown by
  \citet{2002MNRAS.335..487M}.  The AMD of bright galaxies in
  \citet{2002MNRAS.335..487M}, which are effected little by feedback ,
  seems to have $\alpha \sim 1$ ( no dip in AMD for small s),
  consistent with AM profile of a galaxy with a bulge which accounts
  for the low AM material. It is no way similar to exponential disk
  with flat rotation curves as claimed there, because that would imply
  an $\alpha=2$ and will be reflected as a prominent dip.
\end{enumerate}

\acknowledgements We are grateful to James Bullock and Daniel
Eisenstein for stimulating discussions and also Vince Eke for his help
with the initial conditions of the simulations. This work has been
supported by grants from the U.S. National Aeronautics and Space
Administration (NAG 5-10827), the David and Lucile Packard Foundation,
and the Bundesministerium f\"ur Bildung und Forschung (FKZ
05EA2BA1/8).


\begin{thebibliography}{ }

  
\bibitem[Barnes \& Efstathiou(1987)]{1987ApJ...319..575B} Barnes,
  J.~\& Efstathiou, G.\ 1987, \apj, 319, 575
  
\bibitem[Bryan \& Norman(1998)]{1998ApJ...495...80B} Bryan, G.~L.~\&
  Norman, M.~L.\ 1998, \apj, 495, 80
  
\bibitem[Bullock et al.(2001)]{2001ApJ...555..240B} Bullock, J.~S.,
  Dekel, A., Kolatt, T.~S., Kravtsov, A.~V., Klypin, A.~A., Porciani,
  C., \& Primack, J.~R.\ 2001, \apj, 555, 240
  
\bibitem[Bullock et al.(2001)-2]{2001MNRAS.321..559B} Bullock, J.~S.,
  Kolatt, T.~S., Sigad, Y., Somerville, R.~S., Kravtsov, A.~V.,
  Klypin, A.~A., Primack, J.~R., \& Dekel, A.\ 2001, \mnras, 321, 559
  
\bibitem[Chen \& Jing(2002)]{2002MNRAS.336...55C} Chen, D.~N.~\& Jing,
  Y.~P.\ 2002, \mnras, 336, 55
  
\bibitem[Chen, Jing, \& Yoshikaw(2003)]{2003ApJ...597...35C} Chen,
  D.~N., Jing, Y.~P., \& Yoshikaw, K.\ 2003, \apj, 597, 35
  
\bibitem[Dalcanton, Spergel, \& Summers(1997)]{1997ApJ...482..659D}
  Dalcanton, J.~J., Spergel, D.~N., \& Summers, F.~J.\ 1997, \apj,
  482, 659
  
\bibitem[Efstathiou \& Jones(1979)]{1979MNRAS.186..133E} Efstathiou,
  G.~\& Jones, B.~J.~T.\ 1979, \mnras, 186, 133
  
\bibitem[Eke, Navarro, \& Steinmetz(2001)]{2001ApJ...554..114E} Eke,
  V.~R., Navarro, J.~F., \& Steinmetz, M.\ 2001, \apj, 554, 114
  
\bibitem[Fall \& Efstathiou(1980)]{1980MNRAS.193..189F} Fall, S.~M.~\&
  Efstathiou, G.\ 1980, \mnras, 193, 189
  
\bibitem[Maller \& Dekel(2002)]{2002MNRAS.335..487M} Maller, A.~H.~\&
  Dekel, A.\ 2002, \mnras, 335, 487
  
\bibitem[Melchiorri, Bode, Bahcall, \&
  Silk(2003)]{2003ApJ...586L...1M} Melchiorri, A., Bode, P., Bahcall,
  N.~A., \& Silk, J.\ 2003, \apjl, 586, L1
  
\bibitem[Mo, Mao, \& White(1998)]{1998MNRAS.295..319M} Mo, H.~J., Mao,
  S., \& White, S.~D.~M.\ 1998, \mnras, 295, 319
  
\bibitem[Navarro \& Benz(1991)]{1991ApJ...380..320N} Navarro, J.~F.~\&
  Benz, W.\ 1991, \apj, 380, 320
  
\bibitem[Navarro \& Steinmetz(1997)]{1997ApJ...478...13N} Navarro,
  J.~F.~\& Steinmetz, M.\ 1997, \apj, 478, 13

  
\bibitem[Navarro \& White(1994)]{1994MNRAS.267..401N} Navarro,
  J.~F.~\& White, S.~D.~M.\ 1994, \mnras, 267, 401

  
\bibitem[Navarro, Frenk, \& White(1996)]{1996ApJ...462..563N} Navarro,
  J.~F., Frenk, C.~S., \& White, S.~D.~M.\ 1996, \apj, 462, 563
  
\bibitem[Navarro, Frenk, \& White(1997)]{1997ApJ...490..493N} Navarro,
  J.~F., Frenk, C.~S., \& White, S.~D.~M.\ 1997, \apj, 490, 493
  
\bibitem[Peebles(1969)]{1969ApJ...155..393P} Peebles, P.~J.~E.\ 1969,
  \apj, 155, 393
  
\bibitem[Spergel et al.(2003)]{2003ApJS..148..175S} Spergel, D.~N.~et
  al.\ 2003, \apjs, 148, 175
  
\bibitem[Springel, Yoshida, \& White(2001)]{2001NewA....6...79S}
  Springel, V., Yoshida, N., \& White, S.~D.~M.\ 2001, New Astronomy,
  6, 79
  
\bibitem[Steinmetz \& Bartelmann (1995)]{1995MNRAS...272..570B}
  Steinmetz, M. \& Bartelmann, M.\ 1995, \mnras 272, 570
  
\bibitem[Steinmetz \& Navarro(1999)]{1999ApJ...513..555S} Steinmetz,
  M.~\& Navarro, J.~F.\ 1999, \apj, 513, 555

  
\bibitem[van den Bosch et al.(2002)]{2002ApJ...576...21V} van den
  Bosch, F.~C., Abel, T., Croft, R.~A.~C., Hernquist, L., \& White,
  S.~D.~M.\ 2002, \apj, 576, 21

  
\bibitem[van den Bosch(2001)]{2001MNRAS.327.1334V} van den Bosch,
  F.~C.\ 2001, \mnras, 327, 1334
  
\bibitem[van den Bosch, Burkert, \&
  Swaters(2001)]{2001MNRAS.326.1205V} van den Bosch, F.~C., Burkert,
  A., \& Swaters, R.~A.\ 2001, \mnras, 326, 1205
  
\bibitem[White \& Rees(1978)]{1978MNRAS.183..341W} White, S.~D.~M.~\&
  Rees, M.~J.\ 1978, \mnras, 183, 341



\end{thebibliography}
\end{document}